\documentclass[11pt]{article}
\usepackage{amsmath,amssymb,amsfonts,bbm}

\newcommand{\be}{\begin{equation}}
\newcommand{\ee}{\end{equation}}
\newcommand{\bea}{\begin{eqnarray}}
\newcommand{\eea}{\end{eqnarray}}
\newcommand{\ba}{\begin{array}}
\newcommand{\ea}{\end{array}}

\newcommand{\htwo}{h_{2,1}}
\newcommand{\M}{\mathcal{M}}
\newcommand{\N}{\mathcal{N}}
\newcommand{\D}{\mathcal{D}}
\newcommand{\K}{\mathcal{K}}
\newcommand{\F}{\mathcal{F}}
\newcommand{\A}{\mathcal{A}}

\long\def\symbolfootnote[#1]#2{\begingroup%
\def\thefootnote{\fnsymbol{footnote}}\footnote[#1]{#2}\endgroup}

\begin{document}

\thispagestyle{empty}\vspace{40pt}

\hfill{}

\vspace{128pt}

\begin{center}
    \textbf{\Large The many symmetries of Calabi-Yau compactifications}\\
    \vspace{40pt}

    Moataz H. Emam\symbolfootnote[1]{Electronic address: {\tt moataz.emam@cortland.edu}}

    \vspace{12pt}   \textit{Department of Physics}\\
                    \textit{SUNY College at Cortland}\\
                    \textit{Cortland, NY 13045, USA}\\
\end{center}

\vspace{40pt}

\begin{abstract}

We review the major mathematical concepts involved in the dimensional reduction of $D=11$ $\N=1$ supergravity theory over a Calabi-Yau manifold with non-trivial complex structure moduli resulting in ungauged $D=5$ $\N=2$ supergravity theory with hypermultiplets. This last has a particularly rich structure with many underlying geometries. We reproduce the entire calculation and particularly emphasize its symplectic symmetry and how that arises from the topology of the underlying subspace. The review is intended to fill in a specific gap in the literature with the hope that it would be useful to both the beginner and the expert alike.

\end{abstract}

\newpage


\tableofcontents

\vspace{15pt}

\pagebreak

\section{Introduction}

It has long been hoped that the use of Kaluza-Klein techniques to dimensionally reduce string/ supergravity (SUGRA) theories will eventually lead to a physically acceptable four dimensional representation of our universe, \emph{i.e.} the standard model plus gravity. Unfortunately the number of possible ways of doing so turns out to be (almost) unbelievably high. In fact, the figure $10^{500}$ is often quoted. It is further speculated that a specific choice of vacuum (\emph{i.e.} the choice of compactification subspace, its topological parameters etc) would pick the correct four dimensional structure by some sort of physical `natural selection' mechanism. This problem of the so-called `String Theory Landscape' \cite{Susskind:2005bd, Freivogel:2005vv, Freivogel:2004rd, Susskind:2004uv, Susskind:2003kw} is currently the major obstacle in our understanding of string theory as the most promising theory of everything, and is, in fact, the main argument raised by the theory's critics (and a good argument no doubt) \cite{Smolin:2006pe, Woit:2006js}\footnote{Also see \cite{Emam:2008js} for a counter argument.}. It then becomes of paramount importance to understand the mathematical techniques of dimensional reduction. These normally involve understanding the geometries and topologies of manifolds with special holonomy, as well as specific types of complex manifolds that arise as a consequence of the dimensional reduction. There is of course a huge literature available on these topics, including discussions written by both physicists and mathematicians. However, there does not seem to be a single source that would act as a tutorial to the beginner, discussing the calculation from the most basic of definitions all the way to completion. This is further complicated by the lack of a unified notation for the various topics. As such it is quite hard for the beginner to follow and reproduce the results in full. This review intends to fill this particular gap. Our choice of specific calculation to reproduce is that of the reduction of eleven dimensional $\N=1$ supergravity over a Calabi-Yau 3-fold with non-trivial complex structure moduli. This leads to a five dimensional $\N=2$ theory with a matter sector comprised of an arbitrary number of scalar fields (and their supersymmetric partners); the so-called hypermultiplets. This theory is rarely discussed in the literature, particularly not in the form we review here; another gap we intend to fill.

The study of $\N=2$ supergravity theories in general has gained
interest in recent years for a variety of reasons. For example,
$\N=2$ branes are particularly relevant to the conjectured
equivalence between string theory on anti-de Sitter space and
certain superconformal gauge theories living on the boundary of
the space (the AdS/CFT duality) \cite{Maldacena:2001uc}. Also
interesting is that many results were found to involve the
so-called attractor mechanism (\emph{e.g.}
\cite{hep-th/9508072,hep-th/9602111,hep-th/9602136}); the study of
which developed very rapidly with many intriguing outcomes
(\emph{e.g.} \cite{hep-th/0506177,hep-th/0507096,hep-th/0508042}).
From the point of view of dimensional reduction, many $D=4,5$ results were
shown to be related to higher dimensional ones via wrapping over
specific cycles of manifolds with special holonomy. For example,
M-branes wrapping K\"{a}hler cycles of a Calabi-Yau
(CY) 3-fold \cite{Cho:2000hg} dimensionally reduce to black holes
and strings coupled to the vector multiplets of five dimensional
$\N=2$ supergravity \cite{Kastor:2003jy}, while M-branes wrapping
special Lagrangian cycles reduce to configurations
carrying charge under the hypermultiplet scalars
\cite{Martelli:2003ki,Fayyazuddin:2005as,Emam:2005bh,Emam:2006sr,Emam:2007qa}.

In reviewing the literature, one notices that most studies in
$\N=2$ SUGRA in any number of dimensions specifically address the
vector multiplets sector; setting the hypermultiplets to zero.
This is largely due to the fact that the standard representation
of the hypermultiplet scalars as coordinates on a quaternionic
manifold is somewhat hard to deal with. It has been shown,
however, that certain duality maps relate the target space of a
given higher dimensional fields' sector to that of a lower
dimensional one \cite{Ferrara:1989ik}. Particularly relevant to
this review is the so-called c-map which relates the quaternionic
structure of the $D=5$ hypermultiplets to the more well-understood
special geometric structure of the $D=4$ vector multiplets. This
means that one can recast the $D=5$ hypermultiplet fields into a
form that makes full use of the methods of special geometry. This
was done in \cite{Gutperle:2000ve} and applied in the
same reference as well as in \cite{Emam:2005bh} and others. Using
this method, finding solutions representing the five dimensional
hypermultiplet fields often means coming up with ans\"{a}tze that
have special geometric form. This can be, and has been, done by
building on the considerable $D=4$ vector multiplets literature,
and in most cases the solutions are remarkably similar. For
example, $D=5$ hypermultiplet couplings to 2-branes and instantons (\cite{Emam:2005bh} and \cite{Gutperle:2000ve}) lead to the same
type of attractor equations found for the vector multiplets
coupled to $D=4$ black holes (\emph{e.g.}
\cite{Behrndt:1997fq,Sabra:1997dh,Behrndt:1997ny,Behrndt:1998eq}).

Furthermore, it has long been known that quaternionic and
special K\"{a}hler geometries contain symplectic isometries and
that the hypermultiplets action (with or without gravity) is in
fact symplectically invariant\footnote{This being a
straightforward generalization of the ordinary Maxwell dualities
first discussed from within the context of supergravity in
\cite{Ferrara:1976iq}, but generally known for ordinary
electrodynamics since 1925 \cite{Rainich}.}. The exploitation of this particular property was recently proposed as a method of constructing solutions to the theory \cite{Emam:2009xj}. We include this in our review and emphasize the origin of the symplectic structure of the theory from the topology of the subspace. The discussion is not
intended to be exhaustive, rather enough information is presented
to achieve an overall, hopefully intuitive, understanding of the
process as well as provide a hands-on first reading. Some of this review is based on
the author's \cite{Emam:2004nc}. Further detail may be sought out
in the given cited texts.

The review is structured in the following way: Starting from basic principles, section
\ref{manifoldsreview} discusses the various types of complex
manifolds needed in the rest of the review. Section
\ref{SKGandSp} focuses on special K\"{a}hler geometry with
particular emphasis on its symplectic structure. Section \ref{dimensionalreduction} presents
the details of the dimensional reduction of $D=11$ SUGRA over a
Calabi-Yau 3-fold with non-trivial complex structure moduli. For easy reference we include an appendix on the basics of the language of differential forms on manifolds.

\section{Manifolds; from Riemann to Yau\label{manifoldsreview}}

We review the various classes of complex manifolds we will need.
Starting with elementary definitions, we write down the different
properties with minimal mathematics. The discussion is by no means
exhaustive, but enough material is reviewed in preparation for a, hopefully,
intuitive understanding of the process of dimensional reduction. Where appropriate, we use the
language of differential forms as defined in the appendix.

\subsection{Complex and K\"{a}hler manifolds}

We define the notion of a real $n$-dimensional \textbf{manifold}
$\M$ as a set of points that behaves locally like
$\mathbb{R}^{n}$, such that $n$ real parameters ($x^1, \ldots,
x^\alpha, \ldots, x^{n}$) are coordinates on $\M$
\cite{Wald:1984rg, Nakahara:2003nw}. Similarly, a complex
$k$-dimensional manifold may be defined as a set of points that
behaves locally like $\mathbb{C}^k$, where $\left\{ {n,k \in
\mathbb{Z}} \right\}$. A \textbf{Riemannian manifold} is a
manifold on which a smooth symmetric positive-definite metric
tensor $g_{\mu \nu } \left( {x^\alpha  } \right)$ can be defined,
describing a line element on the manifold $ds^2  = g_{\mu \nu }
dx^\mu dx^\nu$. A manifold is called \textbf{Lorentzian} if its
metric has Lorentzian signature\footnote{Which we take to be
$\left( { - + + \cdots + } \right)$ throughout.}, \emph{i.e.}
behaves locally like $\mathbb{R}^{1,n-1}$. A Levi-Civita
connection (\emph{i.e.} metric-compatible) may be chosen, leading
to the usual expressions for the Christoffel symbols, the Riemann
and Ricci tensors and the Ricci scalar:
\begin{eqnarray}
    \Gamma _{\mu \nu }^\lambda   &=& \frac{1}{2}g^{\lambda \kappa } \left[ {\left(
    {\partial_\mu  g_{\nu \kappa } } \right) + \left( {\partial _\nu  g_{\mu \kappa } }
    \right) - \left( {\partial _\kappa  g_{\mu \nu } } \right)} \right] \nonumber \\
    R_{\mu \rho \nu }^{\,\,\,\,\,\,\,\,\,\,\,\sigma }  &=& \left( {\partial _\rho  \Gamma _{\mu \nu }^\sigma  } \right) - \left( {\partial _\mu  \Gamma _{\rho \nu }^\sigma  } \right) + \Gamma _{\mu \nu }^\alpha  \Gamma _{\alpha \rho }^\sigma   - \Gamma _{\rho \nu }^\alpha  \Gamma _{\alpha \mu }^\sigma\nonumber \\
    R_{\mu \nu }  &=& R_{\mu \rho \nu }^{\,\,\,\,\,\,\,\,\,\,\,\rho }  = \left( {\partial _\rho  \Gamma _{\mu \nu }^\rho  } \right) - \left( {\partial _\mu  \Gamma _{\rho \nu }^\rho  } \right) + \Gamma _{\mu \nu }^\alpha  \Gamma _{\alpha \rho }^\rho   - \Gamma _{\rho \nu }^\alpha  \Gamma _{\alpha \mu }^\rho\nonumber\\
    R &=& g^{\mu \nu } R_{\mu \nu } =R_\mu ^\mu.
\end{eqnarray}

If one now considers manifolds with even dimensions, \emph{i.e.}
$n=2k$, then one can, at least locally, `complexify' $\M$ by
pairing $x^\alpha$ as follows (summation convention not used):
\begin{eqnarray}\label{complex}
    w^\alpha   &=& x^\alpha   + \tau _{\alpha  + k}^\alpha  x^{\alpha  + k} \equiv w^m,\quad \quad m = 1, \ldots ,k  \nonumber\\
    \bar w^\alpha   &=& x^\alpha   + \bar \tau _{\alpha  + k}^\alpha  x^{\alpha  + k}  \equiv w^{\bar m },\quad \quad \bar m = {\bar 1}, \ldots
    ,{\bar k}
\end{eqnarray}
where the $\tau$'s are complex parameters that specify a complex
structure on the manifold (more on that later). A general metric
on such a manifold is then:
\be
    ds^2  = g_{mn} dw^m dw^n  + g_{\bar m\bar n} dw^{\bar m} dw^{\bar n}  + 2g_{m\bar n} dw^m dw^{\bar
    n}.
\ee

Reality of the line element is insured by the conditions
\begin{eqnarray}
    g_{mn}&=&g_{\bar m \bar n} \nonumber \\
    g_{m \bar n}&=&g_{n \bar m}.
\end{eqnarray}

A \textbf{Hermitian manifold} is defined as a complex manifold
where there is a preferred class of coordinate systems such
that\footnote{It can actually be shown that a Hermitian metric may
be constructed on any complex manifold. As such, the term
`Hermitian manifold' is somewhat misleading. Nevertheless,
following the literature, we will continue to use it with the
understanding that we are really talking about a complex
\emph{manifold} with a Hermitian \emph{metric}.}:
\begin{equation}\label{Hermitian}
    g_{mn}=g_{\bar m \bar n}=0.
\end{equation}

The line element reduces to
\begin{equation}\label{}
    ds^2  = 2g_{m\bar n} dw^m dw^{\bar n}.
\end{equation}

On any Hermitian manifold, a \emph{real} 2-form, known as the
\textbf{K\"{a}hler form}, can be defined as a (1,1) form as
follows:
\begin{equation}\label{form}
    K  =  ig_{m\bar n} dw^m  \wedge dw^{\bar n}.
\end{equation}

A \textbf{K\"{a}hler manifold} is a Hermitian manifold whose
K\"{a}hler form is closed, \emph{i.e.}
\be\label{Kahlercondition}
    dK= 0.
\ee

As a closed 2-form, the K\"{a}hler form is a member of a
cohomology class; namely the second De-Rahm class $[K]  \in H^2
\left( \M \right)$. Treating $K$ as a (1,1)-form in our complex
basis, this corresponds to the $H^{1,1}$ Dolbeault class.
Henceforth we will refer to $H^{1,1}$ as the K\"{a}hler class of
the metric. Equation (\ref{Kahlercondition}) leads to the
`curl-free' condition:
\begin{equation}\label{kahler}
    \partial _m g_{n\bar p}  - \partial _n g_{m\bar p} = 0,
\end{equation}
which may equivalently be used as the definition of a K\"{a}hler
manifold\footnote{Note that not all complex manifolds admit
K\"{a}hler metrics.}. This implies that locally the K\"{a}hler
metric can be determined in terms of a real scalar function, known
as the \textbf{K\"{a}hler potential} ${\K}(w,\bar w)$. In other
words (\ref{kahler}) is solved by:
\begin{equation}
    g_{m\bar n}  = \partial _m \partial _{\bar n} \K\quad  \to\quad K  = i\left( {\partial _m \partial _{\bar n} \K} \right)dw^m  \wedge dw^{\bar n}.
\end{equation}

Obviously, the metric is invariant under changes of the K\"{a}hler
potential of the form $\K\left( {w,\bar w} \right) \to \K\left(
{w,\bar w} \right) + f_1\left( w \right) + f_2\left( {\bar w}
\right)$, known as the \textbf{K\"{a}hler gauge transformations}.
It follows then that if two K\"{a}hler metrics on $\M$ belong to
the same K\"{a}hler class, then they can differ \emph{only} by a
K\"{a}hler transformation.

From a computational point of view, the condition
(\ref{Kahlercondition}), or equivalently (\ref{kahler}),
simplifies the properties of the manifold considerably, for
example one finds that
\bea
    \Gamma _{mn}^r  &=& g^{r\bar p} \left( {\partial _m g_{n\bar p} }
    \right),\nonumber\\  \Gamma _{\bar m\bar n}^{\bar r}  &=& g^{p\bar r} \left( {\partial _{\bar m} g_{\bar np} }
    \right),\label{LeviCivitaonKahler}
\eea
are the only non-vanishing Christoffel symbols, indicating that
parallel transport does not mix the holomorphic with the
antiholomorphic components of a vector. Also the non-vanishing
components of the Ricci tensor are found to be
\bea
    R_{m\bar n}  &=& \partial _m \partial _{\bar n} \ln g, \nonumber\\ {\rm where\;\;} g &=& \det  {g_{m\bar n} }.\label{ricci}
\eea

\subsection{Issues of global importance}

Technically, the assumption that any real $2k$-dimensional
manifold $\M$ can be made into a complex manifold is only valid
locally. Global considerations must be included in order to
properly decide if a given manifold is truly complex everywhere. A
key element to such considerations is the so-called
\textbf{complex structure} of the manifold. Intuitively, it is
nothing more than the formalization of multiplication by $i$
smoothly over the manifold, \emph{i.e.} an operation on
geometrical objects whose square is negative the identity. A
tensor $J$ on $\M$ is called an \emph{almost} complex structure if
it satisfies the condition:
\begin{eqnarray}
    J^2 \sim - \mathbbm{1}\quad :\quad
    J_\mu ^\rho  \left( x \right)J_\rho ^\nu  \left( x \right) =  -
    \delta _\mu ^\nu,
\end{eqnarray}
where the $2k$ real Greek indices break into $\left(m,\bar
n\right)$ as before. In components, $J$ is related to the
K\"{a}hler form by:
\be
    K_{\mu \nu }  = g_{\nu \rho } J_\mu ^\rho,\label{complexstructure}
\ee
and is also related to the complex parameters in (\ref{complex}).
For example, one common choice is
\be
    \tau _\nu ^\mu   = iJ_\nu ^\mu  ,\quad \quad \bar \tau _\nu ^\mu   =  - iJ_\nu
    ^\mu.
\ee

Now if a manifold $\M$ has a smooth almost complex structure, it
is called an almost complex manifold. An almost complex structure
becomes a complex structure when its so-called \textbf{Nijenhuis
tensor}
\begin{equation}
    N_{\mu \nu }^\rho   = J_\mu ^\alpha  \left[ {\left( {\partial
    _\alpha  J_\nu ^\rho  } \right) - \left( {\partial _\nu  J_\alpha
    ^\rho  } \right)} \right] - J_\nu ^\alpha  \left[ {\left(
    {\partial _\alpha  J_\mu ^\rho  } \right) - \left( {\partial _\mu
    J_\alpha ^\rho  } \right)} \right]
\end{equation}
vanishes everywhere. This condition is achieved by demanding that
different complex structures on a manifold smoothly patch
together\footnote{A given real manifold can admit many complex
structures.}. So, any $2k$-dimensional real manifold is locally
complex (almost complex manifold), but only globally so (complex
manifold) when it admits a complex structure with vanishing
Nijenhuis tensor. This is analogous to the concept that any
Riemannian manifold is locally flat, but only globally so when the
Riemann tensor vanishes everywhere. Consequently one may speak of
`almost Hermitian manifolds', `almost K\"{a}hler manifolds' and so
on.

Another point of global importance is the question of holonomy
groups on a K\"{a}hler manifold \cite{Joyce}. Consider a vector
$V^\mu$ on an $n$-fold and parallel transport it around a closed
loop, generally the vector will not return to itself, but rather
rotated by an element of $GL(n,\mathbb{R})$. The subset of
$GL(n,\mathbb{R})$ defined in this way forms the \textbf{holonomy
group} of the manifold. The \textbf{restricted holonomy group}
would be the subset defined by paths which may be smoothly shrunk
to a point (contractable loops). The classification of the
restricted holonomy groups of all Riemannian manifolds has been
performed by Berger \cite{Berger}, which we list for completeness:

\begin{description}
    \item[Berger's theorem:] Suppose $\mathcal{M}$ is a simply-connected
    manifold of dimension $n$, and that $g$ is a Riemannian metric on
    $\mathcal{M}$, then \emph{exactly} seven restricted, or special, holonomy cases are possible:
        \begin{enumerate}
            \item Generic Riemannian manifolds, $Hol(g) = SO(n)$.
            \item K\"{a}hler manifolds, where $n = 2k$ with $k \ge
            2$ and $Hol(g) = U(k) \subset SO(2k)$.
            \item Calabi-Yau manifolds, where $n = 2k$ with $k \ge
            2$ and $Hol(g) = SU(k) \subset SO(2k)$. These are also necessarily Ricci-flat (Yau's theorem).
            \item HyperK\"{a}hler manifolds, where $n = 4k$ with $k \ge
            2$ and $Hol(g) = Sp(k) \subset SO(4k)$.
            \item Quaternionic K\"{a}hler manifolds, where $n = 4k$ with $k \ge
            2$ and $Hol(g) = Sp(k)\otimes Sp(1) \subset SO(4k)$.
            \item Manifolds with $n = 7$ and $Hol(g) = G_2 \subset SO(7)$.
            \item Manifolds with $n = 8$ and $Hol(g) = Spin(7) \subset
            SO(8)$. The groups $G_2$ and $Spin(7)$ are exceptional holonomy groups.
        \end{enumerate}
\end{description}

We can categorize the holonomy groups in Berger's list as follows:
\begin{itemize}
    \item The K\"{a}hler holonomy groups: $U(k)$, $SU(k)$ and
    $Sp(k)$. Any Riemannian manifold with one of these is
    necessarily K\"{a}hler.
    \item The Ricci-flat holonomy groups: $SU(k)$, $Sp(k)$, $G_2$
    and $Spin(7)$. Any metric with one of these is necessarily
    Ricci-flat.
    \item The exceptional holonomy groups: $G_2$ and $Spin(7)$.
    So-called because they have properties fundamentally different
    from the others.
\end{itemize}

The Berger list may also be understood in terms of the four
division algebras in the following way: It is well-known that one
can define exactly four algebras where, for two quantities
$\mathcal{Z}_1$ and $\mathcal{Z}_2$, the property $\left|
{\mathcal{Z}_1 \mathcal{Z}_2 } \right| = \left| {\mathcal{Z}_1 }
\right|\left| {\mathcal{Z}_2 } \right|$ is satisfied. These are
the real numbers $\mathbb{R}$, the complex numbers $\mathbb{C}$,
the quaternions $\mathbb{H}$ and the octonions, or Cayley numbers,
$\mathbb{O}$. The Berger list fits into this classification by
noting that:
\begin{itemize}
  \item $SO(n)$ is a group of automorphisms of ${\mathbb{R}}^n$.
  \item $U(k)$ and $SU(k)$ are groups of automorphisms of ${\mathbb{C}}^k$.
  \item $Sp(k)$ and $Sp(k)\otimes Sp(1)$ are groups of automorphisms of ${\mathbb{H}}^k$.
  \item $G_2$ is a group of automorphisms of ${\mathop{\rm Im}\nolimits} \mathbb{O} \approx \mathbb{R}^7 $.
  \item $Spin(7)$ is a group of automorphisms of $\mathbb{O} \approx \mathbb{R}^8 $.
\end{itemize}

It is interesting to note that all of the manifolds on Berger's
list have found applications in theoretical physics. In fact, the
$\N=2$ theory we will be discussing makes use of all of them
except the exceptional manifolds.

\subsection{Hodge-K\"{a}hler manifolds}

We recall that given a Riemannian manifold $\mathcal{M}$ endowed
with a metric $g_{\mu \nu}$, one can define the vielbeins $e^{\hat
a}$, the connection 1-form $\omega ^{\hat a\hat b}$ (a.k.a. spin
connection) in the following way:
\begin{eqnarray}
    ds^2  &=& g_{\mu \nu } dx^\mu  dx^\nu   =   \eta _{\hat a\hat c}e^{\hat a} e^{\hat c}  ; \quad e^{\hat a}  = e_{\;\; \mu} ^{\hat a} dx^\mu   \label{bein} \\
    \omega _\mu ^{\;\;\hat a\hat c}  &=& e^{\hat a\lambda } \left[ {\left({\partial _\mu  e_{\;\; \lambda} ^{\hat c} } \right) - \Gamma _{\mu
    \lambda }^\nu  e_{\;\; \nu} ^{\hat c} } \right]; \quad\omega ^{\hat a\hat b}  = \omega_\mu ^{\;\;\hat a\hat b}
    dx^\mu,
\end{eqnarray}
such that the so called \textbf{Cartan structure equations} define
the torsion and curvature
2-forms:
\begin{eqnarray}
    \Im^{\hat a}  &=& de^{\hat a}  + \omega _{\;\; \hat c}^{\hat a}  \wedge e^{\hat
    c}= \frac{1}{2}\Im^{\hat a} _{\;\;\mu \nu } dx^\mu   \wedge dx^\nu  \nonumber \\
    \Re^{\hat a\hat c}  &=& d\omega ^{\hat a\hat c}  + \omega_{\;\;
    \hat  b}^{\hat a}  \wedge \omega ^{\hat b\hat c} = \frac{1}{2}\Re^{\hat a\hat c} _{\;\;\;\mu \nu } dx^\mu   \wedge dx^\nu ,
\end{eqnarray}
where the hated indices are raised and lowered by the flat metric
$\eta_{\hat a\hat c}$ (which may be either Minkowski or Euclidean
depending on the signature of $g_{\mu \nu}$), describing a flat
space tangent to each point on the manifold. They are also
sometimes referred to as `frame' indices, as opposed to the
manifold's `world' indices.

Using this language, one can define a topological quantity known
as the \textbf{total Chern form}\footnote{Pronounced
``\emph{Chen}''.} \cite{hep-th/0102152}, which is a polynomial in
the curvature as follows:
\begin{equation}
    C\left( \Re  \right) = \det \left( {1 + \frac{i}{{2\pi }}\Re
    } \right) = 1 + c_1 \left( \Re  \right) + c_2 \left( \Re
    \right) +  \cdots.
\end{equation}

The terms $c_i$ are the so-called \textbf{Chern classes}. They
belong to topologically distinct cohomology classes. For example:
\bea
 c_0  &=& 1, \nonumber\\
 c_1  &=& \frac{i}{{2\pi }}{\rm Tr}\Re , \nonumber\\
 c_2  &=& \frac{1}{{8\pi ^2 }}\left[ {{\rm Tr}\left( {\Re  \wedge \Re } \right) - {\rm Tr}\Re  \wedge {\rm Tr}\Re } \right],\quad {\rm \emph{etc}}{\rm
 .}
\eea

Furthermore, integrals such as
\begin{equation}
    \int\limits_\M {c_2 \left( \Re  \right)} \quad {\rm and}\quad
    \int\limits_\M {c_1 \left( \Re  \right) \wedge c_1 \left(
    \Re  \right)} \nonumber
\end{equation}
are topologically invariant \emph{integers}, known as the
\textbf{Chern numbers}.

The Chern classes are widely used in classifying invariant
quantities in classical field theory\footnote{For example, the
vector fields of ordinary $U(1)$ Maxwell and $SU(2)$ Yang-Mills
theories can be treated as fiber bundles on spacetime manifolds
where the Chern classes reduce to the special case of the
so-called \textbf{Pontrjagin classes}. The first and second such
classes represent the ordinary field energy density and Poynting
vector. Other such classes are particularly useful in the
topological classification of magnetic monopoles.}. They can also
be used to topologically distinguish various types of manifolds.
Given the Ricci tensor $R_{m \bar n}$ of a K\"{a}hler manifold, we
can define the $(1,1)$ Ricci form
\begin{equation}
    \mathcal{\breve{R}} = R_{m\bar n} dw^m \wedge dw^{\bar n}.\label{RicciForm}
\end{equation}

Since the Ricci form is necessarily closed; $d\mathcal{\breve
R}=0$, then it defines an equivalence class in $H^{1,1}$. The
first Chern class is simply:
\begin{equation}
    c_1  = \frac{i}{{2\pi }}\mathcal{\breve{R}}.
\end{equation}

Now consider a line bundle $\mathcal{L}$ over a K\"{a}hler
manifold $\M$. By definition, this is a holomorphic vector bundle
of rank one\footnote{In a more pedestrian physics language, a
vector bundle on $\M$ is a vector field living on a space or
spacetime manifold $\M$.}. The first Chern class is the only one
that exists for such a bundle. In terms of some Hermitian fibre
metric $h$ on $\mathcal{L}$, and using (\ref{ricci}), this is
clearly:
\be
    c_1 \left( \mathcal{L} \right) = \frac{i}{{2\pi }}\left( {\partial \bar \partial \ln h}
    \right),
\ee
where $\partial  \equiv dw^n \partial _n $ and $\bar \partial
\equiv dw^{\bar n} \partial _{\bar n} $. Since $\mathcal{L}$ is a
line bundle, its connection (Christoffel symbol) is a 1-form
defined by $h$ as follows
\begin{equation}
    \vartheta   = \partial \ln h  ,\quad \bar \vartheta = \bar\partial \ln
    h.
\end{equation}

Also, it is known that there exists a correspondence between line
bundles and $U(1)$ bundles. At the level of connections this
reduces to
\begin{equation}\label{U1connection}
    U(1)\; {\rm connection} \equiv \mathcal{P} = {\mathop{\rm Im}\nolimits} \vartheta = -
    \frac{i}{2}\left( {\vartheta  - \bar \vartheta } \right).
\end{equation}

Now, if $c_1(\mathcal{L})$ happens to equal the cohomology class
of the manifold's K\"{a}hler form (as may be required by the
constraints of supersymmetry for example);
\begin{equation}\label{hodge-k}
    c_1 \left( \mathcal{L} \right) = \left[ K \right],
\end{equation}
then we call this a \textbf{Hodge-K\"{a}hler} manifold
\cite{hep-th/9605032}. An equivalent definition is that the
exponential of the K\"{a}hler potential of the manifold is equal
to the metric of the line bundle. So, a K\"{a}hler manifold with
$\mathcal{L}$ is Hodge-K\"{a}hler if:
\begin{equation}\label{hodge}
    h\left( {w,\bar w} \right) = e^{\K\left( {w,\bar w}
    \right)},
\end{equation}
which enables us to write:
\bea
    c_1 \left( \mathcal{L} \right) &=& \frac{i}{{2\pi }}   \left( \partial \bar \partial
    \K\right)\nonumber\\
    \vartheta  &=& \left( {\partial \K} \right) ,\quad \bar \vartheta  = \left( {\bar \partial
    \K} \right),\nonumber\\
    \mathcal{P} &=&  - \frac{i}{2}\left[ {\left( {\partial  \K} \right)  -
    \left( {\bar\partial  \K} \right) }
    \right].\label{U1connection2}
\eea

A $U(1)$ covariant derivative can then be constructed as follows:
\be
    \nabla  = d + ip\mathcal{P},
\ee
or in components:
\be
    \nabla_n  = \partial_n  + \frac{p}{2}\left( {\partial_n\K} \right),\quad \quad \nabla_{\bar n}  = \partial_{\bar n}  - \frac{p}{2}\left( { \partial_{\bar n} \K}
    \right),\label{U1 cov der}
\ee
where the so-called \textbf{K\"{a}hler weight} $p$ is a constant
determined by the choice of basis. For example, a quantity $W$ on
$\M$ is said to have K\"{a}hler weights $\left(p, \bar p\right)$
if
\bea
    \nabla _n W &=& \left[ {\partial _n  + \frac{p}{2}\left( {\partial _n \K} \right)} \right]W,\quad \quad \nabla _{\bar n} W = \left[ {\partial _{\bar n}  - \frac{p}{2}\left( {\partial _{\bar n} \K} \right)} \right]W \nonumber\\
    \nabla _n \bar W &=& \left[ {\partial _n  + \frac{{\bar p}}{2}\left( {\partial _n \K} \right)} \right]\bar W,\quad \quad \nabla _{\bar n} \bar W = \left[ {\partial _{\bar n}  - \frac{{\bar p}}{2}\left( {\partial _{\bar n} \K} \right)} \right]\bar
    W.
\eea

Furthermore, if $W$ transforms as a tensor on $\M$, then in
addition to coupling to $\mathcal{L}$ via the $U(1)$ connection it
also couples to the metric on $\M$ via the ordinary Levi-Civita
connection (\ref{LeviCivitaonKahler}). The covariant derivative
would then contain both. For example, if $W$ is a vector then:
\be
    \D_n W_m  = \nabla _n W_m  - \Gamma _{nm}^r
    W_r,\quad\quad\quad\quad
    \D_n W_{\bar m}  = \nabla _n W_{\bar m},\label{CovariantLeviKahler}
\ee
and so on for higher rank tensors.

\subsection{Special K\"{a}hler manifolds; a first look \label{SKM First}}

Strictly speaking, there are two types of Special K\"{a}hler
manifolds; dubbed `local' and `rigid'. The former describes the
fields of a locally supersymmetric theory, \emph{i.e.} a
supergravity theory, while the latter pertains to fields in a flat
background. Since our interest is supergravity, we will only
discuss the local type. Sometimes, this type of manifolds is
referred to simply as `special manifolds' and the geometry that
describes it is known as `special K\"{a}hler geometry' or just
`special geometry'.

A \textbf{special K\"{a}hler manifold} of the local type is
defined as a Hodge-K\"{a}hler manifold that admits a completely
symmetric and covariantly holomorphic tensor $C_{mnp}$ and its
antiholomorphic conjugate $C_{\bar m\bar n\bar p}$ such that the
following restriction on the curvature is true:
\be
    R_{\bar mn\bar pq}  = g_{n\bar p} g_{q\bar m}  + g_{q\bar p} g_{n\bar m}  - C_{rqn} C_{\bar s\bar m\bar p} g^{r\bar
    s}.\label{specialcurvature}
\ee

This is generally referred to in the literature as the special
K\"{a}hler geometry (\textbf{SKG}) constraint. The consequences to
(\ref{specialcurvature}) can be calculated, and a large literature
exists on this. However a second, alternative but completely
analogous, definition of special K\"{a}hler manifolds is more
frequently used in the physics literature. It relies heavily on
the symplectic symmetry of special manifolds, a topic of
particular interest to us, so we will develop this concept in a
bit more detail later.

\subsection{Calabi-Yau manifolds}

In 1954 Calabi proposed the following conjecture: If $\mathcal{M}$
is a complex manifold with a K\"{a}hler metric and vanishing first
Chern class, then there exists a unique Ricci flat metric for each
K\"{a}hler class on $\mathcal{M}$. In 1976, Calabi's conjecture
was proven by Yau, also showing that a Ricci flat metric
necessarily has $SU(k)$ holonomy; $k$ being the number of complex
dimensions of $\M$. We then define \textbf{Calabi-Yau manifolds}
(CY) as K\"{a}hler manifolds with Ricci flat ($c_1=0$) metrics.

From its general properties, it turns out that a large number of
different CY manifolds exist. It also turns out that defining them
explicitly is a difficult task. Indeed, very few explicit CY
metrics have ever been written down, and no non-trivial compact
ones are known. However, the properties of CY manifolds make it
possible to work with them without explicit knowledge of the
metric, as far as string/supergravity theory compactifications are
concerned. Yau's theorem in particular guarantees the existence of
a metric. On the other hand, this does impose restrictions on how
far one can specify solutions in the reduced theory, since
generally the solutions will be dependent on the unknown metric of
the subspace, as we will see later. We will restrict ourselves to
six real-dimensional CY manifolds admitting $SU(3)$ holonomy,
since this is the type of interest to string theory in general and
to this work in particular.

The importance of this class of manifolds to physics lies in the
fact that they admit covariantly constant spinors. As a
consequence, it can be shown \cite{Candelas:1985hv} that string
theory compactifications over CY 3-folds preserve some
supersymmetry (also see \cite{hep-th/9506150} and the references
therein). Such compactifications have indeed yielded rich and
physically interesting theories in lower dimensions. Specifically,
the fields in the compactified theory correspond to the parameters
that describe possible deformations of the CY 3-fold. This
parameters' space factorizes, at least locally, into a product
manifold ${\mathcal{M}}_C \otimes {\mathcal{M}}_K$, with
${\mathcal{M}}_C$ being the manifold of the complex structure moduli and
${\mathcal{M}}_K$ being a complexification of the parameters of
the K\"{a}hler class. These so-called \textbf{moduli spaces} turn out to
belong to the category of special K\"{a}hler manifolds. In
addition, there exists a symmetry in the structures of
${\mathcal{M}}_C$ and ${\mathcal{M}}_K$ which lends support to the
so-called mirror symmetry hypothesis of CY 3-folds
\cite{Ferrara:1991na}.

In terms of homology groups, Calabi-Yau 3-folds admit a
non-trivial $H^3$ that can be Hodge-decomposed as follows:
\be
    H^3  = H^{3,0}  \oplus H^{2,1}  \oplus H^{1,2}  \oplus
    H^{0,3}.\label{homologydecomposition}
\ee

The full homology structure is summed up by the so-called Hodge
diamond:
\begin{equation}\label{diamond}
    \begin{array}{*{20}c}
       {} & {} & {} & 1 & {} & {} & {}  \\
       {} & {} & 0 & {} & 0 & {} & {}  \\
       {} & 0 & {} & {h_{1,1} } & {} & 0 & {}  \\
       1 & {} & {h_{1,2} } & {} & {h_{2,1} } & {} & 1  \\
       {} & 0 & {} & {h_{1,1} } & {} & 0 & {}  \\
       {} & {} & 0 & {} & 0 & {} & {}  \\
       {} & {} & {} & 1 & {} & {} & {}
    \end{array}
\end{equation}
where the Hodge numbers $h$ are the dimensions of the respective
homology/cohomology groups the manifold admits\footnote{The
equivalent to the Betti numbers for a real manifold.}, so
(\ref{diamond}) tells us that CY 3-folds have a single (3,0)
cohomology form; $h_{3,0}=\dim \left( {H^{3,0} } \right) = 1$,
which we will call $\Omega$ (the holomorphic volume form) and an
arbitrary number of (1,1) and (2,1) forms determined by the
corresponding $h$'s\footnote{Whose values depend on the
particular choice of CY manifold.}. The Hodge number $h_{2,1}$
determines the dimensions of ${\mathcal{M}}_C$, while $h_{1,1}$
determines the dimensions of ${\mathcal{M}}_K$. The pair
($\mathcal{M},K$), where the K\"{a}hler form $K$ of $\mathcal{M}$
is defined by (\ref{form}), can be deformed by either deforming
the complex structure of $\mathcal{M}$ or by deforming the
K\"{a}hler form $K$ (or both). The space of complex structure moduli ${\mathcal{M}}_C$, which we will explore in detail in the following section, geometrically corresponds to what is known as a \textbf{special Lagrangian} manifold. In the context of the current discussion, such a manifold is defined as a submanifold $L$ of the Calabi-Yau space, \emph{calibrated} with respect to $Re \Omega$, i.e. the pullback of $Re \Omega$ on $L$ is less than or equal to the \emph{volume} of $L$. A more detailed discussion of either the theory of calibrations or the geometry of special Lagrangian manifolds is found in many sources, for example \cite{Joyce}. In string/SUGRA compactifications,
each of the two possible deformations yields a different set of fields in the
lower dimensional theory. One can interpret this in the following
way: the M-branes of $D=11$ SUGRA may, on dimensional reduction,
wrap over either K\"{a}hler submanifolds (also sometimes termed
cycles) of $\M$ or over special Lagrangian
submanifolds, or both. The branes' tension physically deforms the CY manifold
such that $\delta g_{m \bar n}$ corresponds to the former via
(\ref{form}) while non-vanishing $\left( {\delta g_{mn} ,\delta
g_{\bar m\bar n} } \right)$ correspond to the latter via
(\ref{complexstructure}), leading to the deformation of the
complex structure.

Now, let's look at this in some detail. In view of Yau's theorem,
one may consider the parameter space of CY manifolds to be the
parameter space of Ricci-flat K\"{a}hler metrics. Let $g_{\mu\nu}$
and $g_{\mu\nu}+\delta g_{\mu\nu}$ be two Ricci-flat metrics on
$\M$, \emph{i.e.}
\be
    R_{\mu \nu } \left[ {g_{\mu \nu } } \right] = R_{\mu \nu } \left[ {g_{\mu \nu }  + \delta g_{\mu \nu } } \right] =
    0,
\ee
then, along with the metric compatibility condition $\nabla ^\mu
\delta g_{\mu \nu }  = 0$, this leads to
\be
    \nabla ^2 \delta g_{\mu \nu }  + 2R_{\mu \,\,\,\,\nu }^{\,\,\,\,\rho \,\,\,\,\sigma } \delta g_{\rho \sigma }  =
    0,\label{Lichnerowicz}
\ee
known as the \textbf{Lichnerowicz equation}. In fact, it can be
verified that (\ref{Lichnerowicz}) is satisfied for $\delta
g_{m\bar n} $ and $\left( {\delta g_{mn} ,\delta g_{\bar m\bar n}
} \right)$ \emph{independently} of each other. This is
particularly significant to the separation between $\M_C$ and
$\M_K$ alluded to earlier because the deformation of the (1,1)
forms arise from $\delta g_{m\bar n} $ while that of the (2,1)
forms follows from non-vanishing $\left( {\delta g_{mn} ,\delta
g_{\bar m\bar n} } \right)$. For our purposes, we will only
discuss the case with non-vanishing (2,1) forms deformations. In the
five dimensional context, this corresponds to setting the vector
multiplets sector to zero.

\subsection{The space of complex structure moduli}\label{21forms}

It is important to re-emphasize here that $\M_C$ is the space of
the moduli of the complex structure of $\M$ and is \emph{not} a
physical space. It corresponds to special Lagrangian cycles of the
(physical) CY space $\M$ that are completely specified by
knowledge of the unique $(3,0)$ form $\Omega$ and the arbitrary
number of $(2,1)$ forms, which we will call $\chi$. The way the
forms $\chi$ are linked to the complex structure deformations
$\delta g_{m n}$ and $\delta g_{\bar m\bar n}$ is defined via
$\Omega$ as follows \cite{Candelas:1989bb}:
\begin{equation}
    \delta g_{\bar p\bar r}  =  - \frac{1}{{\left\| \Omega  \right\|^2
    }}{\bar\Omega _{\bar p}} ^{\;\;\, mn} \chi _{i|mn\bar r} \delta z^i; \quad
    \left\| \Omega  \right\|^2  \equiv \frac{1}{{3!}}\Omega _{mnp} \bar
    \Omega ^{mnp},\label{modulideformations}
\end{equation}
with the inverse relation
\begin{equation}\label{link}
    \chi _{i|mn\bar p}  =  - \frac{1}{2}{\Omega _{mn}} ^{\bar r} \left(
    {\frac{{\partial g_{\bar p\bar r} }}{{\partial z^i }}}
    \right); \quad \chi _i  = \frac{1}{2}\chi _{i|mn\bar p} dw^m
    \wedge dw^n  \wedge dw^{\bar p},
\end{equation}
which also defines the parameters, or moduli, of the complex
structure $\left( {z^i :i = 1, \ldots ,h_{2,1} } \right)$. Each
$\chi _i $ defines a (2,1) cohomology class. The important
observation here is that the moduli can be treated as complex
coordinates that define a special K\"{a}hler metric $G_{i\bar j}$
on $\mathcal{M}_C$ as follows:
\begin{equation}\label{Gij}
    V_{CY} G_{i\bar j} \left( {\delta z^i } \right)\left( {\delta z^{\bar j}
    } \right) = \frac{1}{{4 }}\int\limits_{\M} {g^{m\bar n} g^{r\bar
    p} \left( {\delta g_{mr} } \right)\left( {\delta g_{\bar n\bar p}
    } \right)},
\end{equation}
where $V_{CY}$ is the volume of the Calabi-Yau. In differential
geometric notation, this gives
\begin{equation}\label{G}
 G_{i\bar j}  =  - \frac{{\int {\chi _i  \wedge \bar \chi _{\bar j}
    } }}{{\int {\Omega  \wedge \bar \Omega } }} = \partial _i
    \partial _{\bar j} \K=  - \partial _i \partial _{\bar j}
    \ln \left( {i\int {\Omega  \wedge \bar \Omega } } \right)
\end{equation}
which also defines its K\"{a}hler potential $\K$. A particularly
useful theorem, attributed to Kodaira, states the following
relations between $\Omega$ and $\chi$:
\begin{equation}\label{Kodaira}
    \left( {\partial _i \Omega } \right) = k_i \Omega  + \chi
    _i,\quad\quad \left( {\partial _{\bar i} \bar\Omega } \right) = k_{\bar i} \bar\Omega  + \bar\chi _{\bar
    i},
\end{equation}
where the arbitrary coefficients $\left(k_i, k_{\bar i}\right)$
may generally depend on the moduli. A reasonable choice for $k_i$
is in fact
\be
    k_i  \propto \left( {\partial _i \K } \right).
\ee

The following can then be demonstrated:
\bea
    \int\limits_\M {\Omega  \wedge \bar \Omega }  &=&  - ie^{ - \K}  \nonumber\\
    \int\limits_\M {\Omega  \wedge \nabla _i \Omega }  &=& \int\limits_\M {\bar \Omega  \wedge \nabla _{\bar i} \bar \Omega }  = 0 \nonumber\\
    \int\limits_\M {\nabla _i \Omega  \wedge \nabla _{\bar j} \bar \Omega }  &=& iG_{i\bar j} e^{ - \K},\label{Omegarelations}
\eea
where the $U(1)$ K\"{a}hler connection $\nabla$ is defined by
(\ref{U1 cov der}). It is well known that the volume of the CY
3-fold is given by\footnote{Generally for CY $k$-folds where $k\ne
3$, the expression (\ref{Volume}) would have different
normalization coefficients.}
\be\label{Volume}
    {\rm Vol}\left( {\rm \M} \right) = i\int\limits_\M {\Omega  \wedge \bar \Omega
    },
\ee
which means that, using the first equation of
(\ref{Omegarelations}), the K\"{a}hler potential of $\M_C$ is
related to the volume of $\M$ simply by
\be
    {\rm Vol}\left( {\rm \M} \right) = e^{ - \K }.
\ee

The space $\mathcal{M}_C$ of complex structure moduli may also be
described in terms of the periods of the holomorphic 3-form
$\Omega$. Let $\left( {A^I ,B_J } \right)$, where $I,J,K = 0,
\ldots ,h_{2,1} $, be a canonical homology basis for $H^3$ such
that
\bea
    A^I  \cap B_J  &=& \delta _J^I ,\nonumber \\ B_I  \cap A^J  &=&  - \delta _I^J  \nonumber\\
    A^I  \cap A^J  &=&  B_I  \cap B_J  = 0,
\eea
and let $\left( {\alpha _I ,\beta ^J } \right)$ be the dual
cohomology basis forms such that
\begin{eqnarray}
    \int\limits_\M {\alpha _I  \wedge } \beta ^J  &=&  \int\limits_{A^J } {\alpha _I } = \delta
    _I^J,\nonumber \\ \int\limits_\M {\beta ^I  \wedge \alpha _J } &=& \int\limits_{B_J } {\beta ^I } =  -
    \delta _J^I , \nonumber \\
    \int\limits_\M {\alpha _I  \wedge } \alpha _J  &=& \int\limits_\M {\beta ^I  \wedge \beta ^J }  =
    0.
    \label{cohbasis}
\end{eqnarray}

The periods of $\Omega$ are then defined by
\begin{equation}\label{periods}
    Z^I  = \int\limits_{A^I } {\Omega },\quad\quad F_I  = \int\limits_{B_I
    } \Omega.
\end{equation}

Now, it can be shown that, locally in the moduli space, the
complex structure is entirely determined by $Z^I$, so one can
write $F_I = F_I \left( Z^J \right)$. Also, a rescaling $Z^I  \to
\lambda Z^I $, where $\lambda$ is a non-vanishing constant,
corresponds to a rescaling of $\Omega$ that does not change the
complex structure, which implies that the $Z^I$'s are projective
coordinates on $\mathcal{M}_C$. In fact, we can choose a set of
independent `special coordinates' $z$ as follows:
\begin{equation}
    z^I  = \frac{{Z^I }}{{Z^0 }},
\end{equation}
which are identified with the complex structure moduli $z^i$. So,
given the cohomology basis defined above, one can invert
(\ref{periods}) as follows\footnote{The significance of the minus
sign in (\ref{defomega}) will become apparent when we discuss the
symplectic structure behind these expressions.}
\begin{equation}\label{defomega}
    \Omega  = Z^I \alpha _I  - F_I \beta ^I,
\end{equation}
and the K\"{a}hler potential of $\mathcal{M}_C$ becomes
\begin{equation}\label{pot}
    \K =  - \ln \left[ {i\left( {\bar Z^I F_I  - Z^I \bar F_I }
    \right)} \right].
\end{equation}

Some of the ingredients of this structure require the knowledge of
the Hodge duality relations (with respect to $\M$) of the forms
$(\alpha, \beta)$ \cite{hep-th/9508001}:
\begin{eqnarray}
    \star\alpha _I  &=& \left( {\gamma _{IJ}  + \gamma ^{KL} \theta _{IK} \theta _{JL} } \right)\beta ^J  - \gamma ^{KJ} \theta _{KI} \alpha _J  \nonumber\\
    \star\beta ^I  &=& \gamma ^{IK} \theta _{KJ} \beta ^J  - \gamma ^{IJ}
    \alpha _J,\label{formsHodgeduality}
\end{eqnarray}
where $\theta_{IJ}$ and $\gamma_{IJ}$ are real matrices defined by
\begin{eqnarray}
\N_{IJ} &=& \bar F_{IJ} +2 i \frac{{N_{IK} Z^K N_{JL} Z^L
}}{{Z^PN_{PQ} Z^Q }}\nonumber\\
   & =& \theta_{IJ}-i \gamma_{IJ}\label{gammathetadefined}
\end{eqnarray}
where $F_{IJ}  = \partial _I F_J $ (the derivative is with respect
to $Z^I$), $N_{IJ}=Im(F_{IJ})$,
$\gamma^{IJ}\gamma_{JK}=\delta^I_K$ and $\N_{IJ}$ is known as the
\textbf{periods matrix}. It is also possible to demonstrate the
useful relation
\be
    \bar Z^I N_{IJ} Z^J  =  - \frac{1}{2}e^{ - \K }.
\ee

One final remark is that it is sometimes possible to further
define $F_I$ as the derivative with respect to $Z^I$ of a
scalar function $F$ known as the \textbf{prepotential},
\emph{i.e.}:
\be
    F_I  = \partial _I F\quad  \to \quad F_{IJ}  = \partial _I \partial _J
    F.
\ee

This, however, is avoided by most authors in the more recent
literature since it is not always possible to find such a
function. It can also be explicitly shown \cite{Craps:1997gp} that
$F$ is not in general invariant under symplectic transformations. In addition, some physically interesting cases
arise precisely when a prepotential does not exist. We will then
follow convention and make no further mention of the
prepotential.

In conclusion, we note that the crucial observation here is that
the curvature of $\M_C$ calculated via the metric $G_{i \bar j}$
satisfies the special K\"{a}hler constraint
(\ref{specialcurvature}). We will develop this further using the
second definition of special geometry in the following sections.

\subsubsection{A simple example}

To get a more intuitive, as well as visual, understanding of the
subject of moduli spaces (by itself a vast topic), we consider the
simplest example of a Calabi-Yau manifold: the ordinary torus
$T^2$ \cite{hep-th/0203247}. In this case, there are two real
periodic degrees of freedom $x$ and $y$, such that:
\begin{equation}
    x = x + R_1  , \quad y = y + R_2,
\end{equation}
corresponding to the $H^1$ homology cycles $A$ and $B$
respectively. The cohomology basis forms would then be:
\begin{eqnarray}
    \alpha  = \frac{{dx}}{{R_1 }} &,&\quad \quad \beta  =  - \frac{{dy}}{{R_2 }}
    \nonumber \\
    \int\limits_A \alpha   =  - \int\limits_B \beta   = 1 &,&\quad \quad
    \int\limits_{T^2 } {\alpha  \wedge \beta }  = 1,
 \end{eqnarray}
such that the volume form is the holomorphic $(1,0)$-form:
\begin{eqnarray}
    \Omega _T  &=& dx + i dy = R_1 \alpha  - iR_2 \beta = Z\alpha  - F\beta  \nonumber \\
    Z &=& \int\limits_A {\Omega _T }  = R_1 \quad  ,\quad \quad F = \int\limits_B
    {\Omega     _T }  = iR_2.
\end{eqnarray}

The metric, the K\"{a}hler form and the ``volume''\footnote{In
this case ``volume'' means the surface area of $T^2$.} of the
torus are then respectively:
\bea
 ds^2  &=& \left\| {\Omega _T } \right\|^2  \\
 K &=& \Omega _T  \wedge \bar \Omega _T  \\
 {\rm Vol} &=& \int\limits_T {\Omega _T  \wedge \bar \Omega _T },
\eea
and the K\"{a}hler potential of $\M_C$ is:
\be
    \K =  - \ln \left( {{\rm Vol}} \right).
\ee

\subsection{A note on quaternionic manifolds\label{Quaternionic Manifolds}}

The subject of \textbf{quaternionic manifolds} (also known as
\textbf{quaternionic K\"{a}hler manifolds}) is part of a larger
class of geometry referred to as \textbf{hyper-K\"{a}hler
geometry}, or simply \textbf{hyper-geometry}, since they are
manifolds that allow for the existence of more than one K\"{a}hler
form. This, in fact, is where the hypermultiplet fields derive
their name from. Just as there are two types of special K\"{a}hler
geometry there are also two types of hyper-geometry, the rigid and
the local. The quaternionic geometry described briefly below is
the local case.

Simply put, a manifold is hyper-K\"{a}hler if it admits an $SU(2)$
bundle that plays the same role here as the $U(1)$ bundle in
special K\"{a}hler manifolds. The manifold is called quaternionic
if the curvature of this bundle is proportional to the manifold's
K\"{a}hler form. The metric of a quaternionic manifold can be
written in the form
\be
    ds^2  = h_{uv}\left(q\right) dq^u dq^v,
\ee
where $(u,v)=1,\ldots, 4n$. Such a manifold admits three complex
structures $J^x$ that satisfy the quaternionic algebra\footnote{As
defined in the appendix, a barred Levi-Civita symbol has the usual
$0$, $1$, $-1$ components.}:
\be
    J^x J^y  =  - \delta ^{xy} \mathbbm{1} +  \bar\varepsilon ^{xyz}J^z  ,\quad \quad x,y,z =
    1,2,3.
\ee

It follows that we can construct three 2-forms known as the
hyperK\"{a}hler forms
\begin{eqnarray}
    K ^x  &=& K _{uv}^x dq^u \wedge dq^v, \\
    K _{uv}^x    &=& h_{uw} \left( {J^x } \right)_v^w,
\end{eqnarray}
generalizing the concept of a K\"{a}hler form. The
hyper-K\"{a}hler forms follow an $SU(2)$ Lie-algebra, in the same
way the ordinary K\"{a}hler form follows a $U(1)$ Lie-algebra.

\section{Special geometry and symplectic covariance}\label{SKGandSp}

In this section, we present the second and most common definition
of special K\"{a}hler geometry. The language we will use relies heavily on the symplectic
structure of special manifolds.

\subsection{Principia symplectica}

Before delving into special geometry proper, we define the
language of symplectic vector spaces and set the notations and
conventions that go with it \cite{Emam:2009xj}. In group theory, the symplectic group
$Sp\left( {2m,\mathbb{F}} \right) \subset GL\left( {2m,\mathbb{F}}
\right)$ is the isometry group of a non-degenerate alternating
bilinear form on a vector space of rank $2m$ over $\mathbb{F}$,
where this last is usually either $\mathbb{R}$ or $\mathbb{C}$,
although other generalizations are possible. For our purposes, we
take $\mathbb{F}=\mathbb{R}$ and $m=\htwo+1$. In other words,
$Sp\left( {2\htwo+2,\mathbb{R}} \right)$ is the group of the real
bilinear matrices
\be
    {\bf \Lambda } = \left[ {\begin{array}{*{20}c}
    {{}^{ {11} }\Lambda _J^I } &  {{}^{ {12} }\Lambda ^{IJ} } \\
    {{}^{ {21} }\Lambda _{IJ} } & {{}^{ {22} }\Lambda _I^J }  \\
    \end{array}} \right] \in Sp\left( {2\htwo+2,\mathbb{R}} \right),\quad \quad {\rm where}\,\,\,I,J = 0, \ldots,\htwo+1
\ee
that leave the totally antisymmetric symplectic matrix:
\be\label{Sp metric}
    {\bf S} = \left[ {\begin{array}{*{20}c}
   0 & \mathbbm{1}  \\
   { - \mathbbm{1}} & {0}  \\
\end{array}} \right] = \left[ {\begin{array}{*{20}c}
   0 & {\delta _I^J }  \\
   { - \delta _J^I } & 0  \\
\end{array}} \right]
\ee
invariant; \emph{i.e.}
\be\label{Sp condition 1}
    {\bf \Lambda }^T {\bf S\Lambda } = {\bf S}\quad\quad\quad\quad\quad\quad{\bf \Lambda }^T {{\bf S}^T{\bf\Lambda} } = {\bf S}^T,
\ee
or equivalently
\bea
    \left[ {{}^{11}\Lambda } \right]^K_I \left[ {^{21} \Lambda } \right]_{JK}  - \left[ {{}^{21}\Lambda } \right]_{IK} \left[ {{}^{11}\Lambda } \right]^K_J &=&0 \nonumber\\
    \left[ {{}^{12}\Lambda } \right]^{IK} \left[ {{}^{22}\Lambda } \right]^J_K  -\left[ {{}^{22}\Lambda } \right]^I_K \left[ {{}^{12}\Lambda } \right]^{JK} &=&0 \nonumber\\
    \left[ {{}^{11}\Lambda } \right]^K_I \left[ {{}^{22}\Lambda } \right]^J_K  - \left[ {{}^{21}\Lambda } \right]_{IK} \left[ {{}^{12}\Lambda } \right]^{JK}  &=&
    \delta^J_I;
\eea
the last of which implies $\left| {\bf \Lambda } \right| =
\mathbbm{1}$. The inverse of ${\bf \Lambda }$ is found to be:
\be
 {\bf \Lambda }^{ - 1}  = {\bf S}^{ - 1} {\bf \Lambda}^T
 {\bf S}=\left[ {\begin{array}{*{20}c}
    {{}^{ {22} }\Lambda _J^I } &  -{{}^{ {12} }\Lambda ^{IJ} } \\
    -{{}^{ {21} }\Lambda _{IJ} } &  {{}^{ {11} }\Lambda _I^J } \\
    \end{array}} \right],\label{Sp condition 2}
\ee
such that, using (\ref{Sp condition 1}), ${\bf \Lambda }^{ - 1}
{\bf \Lambda } = {\bf S}^{ - 1} {\bf \Lambda }^T {\bf S\Lambda } =
{\bf S}^{ - 1} {\bf S} = \mathbbm{1}$ as needed. Also note that
${\bf S}^{ - 1}  = {\bf S}^T  = - {\bf S}$. We adopt the language
that there exists a vector space \textbf{\textit{Sp}} such that
the symplectic matrix $\bf S$ acts as a metric on that space.
Symplectic vectors in \textbf{\textit{Sp}} can be written in a
`ket' notation as follows
\be
    \left| A \right\rangle  = \left( {\begin{array}{*{20}c}
   {a^I }  \\
   {\tilde a_I }  \\
    \end{array}} \right),\quad \left| B \right\rangle  = \left( {\begin{array}{*{20}c}
   {b^I }  \\
   {\tilde b_I }  \\
    \end{array}} \right).
\ee

On the other hand, `bra' vectors defining a space dual to
\textbf{\textit{Sp}} can be found by contraction with the metric
in the usual way, yielding:
\be
    \left\langle A \right| = \left( {{\bf SA}} \right)^T  = {\bf
    A}^T {\bf S}^T  = \begin{array}{*{20}c}
   {\left( {\begin{array}{*{20}c}
   {a^J } & {\tilde a_J }  \\
    \end{array}} \right)}  \\
   {}  \\
    \end{array}\left[ {\begin{array}{*{20}c}
   0 & { - \delta _J^I }  \\
   {\delta _I^J } & 0  \\
    \end{array}} \right] = \begin{array}{*{20}c}
   {\left( {\begin{array}{*{20}c}
   {\tilde a_I } & { - a^I }  \\
    \end{array}} \right)}  \\
   {}  \\
    \end{array},
\ee
such that the inner product on \textbf{\textit{Sp}} is the
`bra(c)ket':
\be
    \left\langle {A}
 \mathrel{\left | {\vphantom {A B}}
 \right. \kern-\nulldelimiterspace}
 {B} \right\rangle  = {\bf A}^T {\bf S}^T {\bf B} = \begin{array}{*{20}c}
   {\left( {\begin{array}{*{20}c}
   {\tilde a_I } & { - a^I }  \\
\end{array}} \right)}  \\
   {}  \\
\end{array}\left( {\begin{array}{*{20}c}
   {b^I }  \\
   {\tilde b_I }  \\
\end{array}} \right) = \tilde a_I b^I  - a^I \tilde b_I  =  - \left\langle {B}
 \mathrel{\left | {\vphantom {B A}}
 \right. \kern-\nulldelimiterspace}
 {A} \right\rangle.\label{Sp inner product}
\ee

In this language, the matrix ${\bf \Lambda }$ can simply be
thought of as a rotation operator in \textbf{\textit{Sp}}. So a
rotated vector is
\be\label{SpRotation}
    \left| {A'} \right\rangle  =  \pm \left|\Lambda A \right\rangle  = \pm {\bf \Lambda A}.
\ee

This is easily shown to preserve the inner product (\ref{Sp inner
product}):
\be
    \left\langle {{A'}}
 \mathrel{\left | {\vphantom {{A'} {B'}}}
 \right. \kern-\nulldelimiterspace}
 {{B'}} \right\rangle  = \left(  \pm  \right)^2 {\bf A}^T {\bf \Lambda }^T {\bf S}^T {\bf \Lambda B} = {\bf A}^T {\bf S}^T {\bf B} = \left\langle {A}
 \mathrel{\left | {\vphantom {A B}}
 \right. \kern-\nulldelimiterspace}
    {B} \right\rangle,
\ee
where (\ref{Sp condition 1}) was used. In fact, one can
\emph{define} (\ref{Sp condition 1}) based on the requirement that
the inner product is preserved. We also define the symplectic invariant
\bea
    \left\langle A \right|\Lambda \left| B \right\rangle  &\equiv& \left\langle {A}
 \mathrel{\left | {\vphantom {A {\Lambda B}}}
 \right. \kern-\nulldelimiterspace}
 {{\Lambda B}} \right\rangle  =  {\bf A}^T {\bf S}^T {\bf \Lambda B}\nonumber\\
 &=&  \left\langle {{A\Lambda ^{ - 1} }}
 \mathrel{\left | {\vphantom {{A\Lambda ^{ - 1} } B}}
 \right. \kern-\nulldelimiterspace}
 {B} \right\rangle=- \left\langle {{ B\Lambda}}
 \mathrel{\left | {\vphantom {{ B\Lambda} A}}
 \right. \kern-\nulldelimiterspace}
 {A} \right\rangle.\label{SpInvariant}
\eea

The matrix $\bf\Lambda$ we will be using in the remainder of the
review has the property
\be
   {}^{22}\Lambda _J^I  =  - {}^{11}\Lambda _J^I \quad  \to \quad {\bf \Lambda }^{ - 1}  =  - {\bf \Lambda
   },\label{LambdaProperty}
\ee
which, via (\ref{SpInvariant}), leads to
\be
    \left\langle A \right|\Lambda \left| B \right\rangle  = \left\langle {A}
 \mathrel{\left | {\vphantom {A {\Lambda B}}}
 \right. \kern-\nulldelimiterspace}
 {{\Lambda B}} \right\rangle  =  - \left\langle {{A\Lambda }}
 \mathrel{\left | {\vphantom {{A\Lambda } B}}
 \right. \kern-\nulldelimiterspace}
     {B} \right\rangle.
\ee

The choice (\ref{LambdaProperty}) is not the only natural one. A
consequence of it is that $\bf \Lambda$ is not symmetric, but
${\bf S \Lambda}$ is. On the other hand an equivalent choice would
be a symmetric $\bf\Lambda$, in which case it would be ${\bf S
\Lambda}$ that satisfies (\ref{LambdaProperty}).

Now consider the algebraic product of the two symplectic scalars
\be\label{inter1}
    \left\langle {A}
 \mathrel{\left | {\vphantom {A B}}
 \right. \kern-\nulldelimiterspace}
 {B} \right\rangle \left\langle {C}
 \mathrel{\left | {\vphantom {C D}}
 \right. \kern-\nulldelimiterspace}
    {D} \right\rangle  = \left( {{\bf A}^T {\bf S}^T {\bf B}} \right)\left( {{\bf C}^T {\bf S}^T {\bf D}} \right).
\ee

The ordinary outer product of matrices is defined by
\be
    {\bf B} \otimes {\bf C}^T  = \left( {\begin{array}{*{20}c}
   {b^I }  \\
   {\tilde b_I }  \\
\end{array}} \right)\begin{array}{*{20}c}
   { \otimes \left( {\begin{array}{*{20}c}
   {c^J } & {\tilde c_J }  \\
\end{array}} \right)}  \\
   {}  \\
\end{array} = \left[ {\begin{array}{*{20}c}
   {b^I c^J } & {b^I \tilde c_J }  \\
   {\tilde b_I c^J } & {\tilde b_I \tilde c_J }  \\
    \end{array}} \right],
\ee
which allows us to rewrite (\ref{inter1}):
\be
    \left\langle {A}
 \mathrel{\left | {\vphantom {A B}}
 \right. \kern-\nulldelimiterspace}
 {B} \right\rangle \left\langle {C}
 \mathrel{\left | {\vphantom {C D}}
 \right. \kern-\nulldelimiterspace}
    {D} \right\rangle  = {\bf A}^T {\bf S}^T \left( {{\bf B} \otimes {\bf C}^T {\bf S}^T } \right){\bf D} = \left\langle A \right|{\bf B} \otimes {\bf C}^T {\bf S}^T \left| D \right\rangle.\label{Inter2}
\ee

Comparing the terms of (\ref{Inter2}), we see that one way a
symplectic outer product can be defined is:
\be
    \left| B \right\rangle \left\langle C \right| = {\bf B} \otimes {\bf C}^T {\bf S}^T  = \left[ {\begin{array}{*{20}c}
   {b^I\tilde c_J} & { - b^I c^J}  \\
   {\tilde b_I\tilde c_J} & { - \tilde b_I c^J}  \\
    \end{array}} \right].\label{Spouterproduct}
\ee

Note that the order of vectors in (\ref{Spouterproduct}) is
important, since generally
\be
    \left| B \right\rangle \left\langle C \right| = \left[ {{\bf
    S}\left| C \right\rangle \left\langle B \right|{\bf S}} \right]^T.
\ee

However, if the outer product $\left| B \right\rangle \left\langle
C \right|$ satisfies the property (\ref{LambdaProperty}),
\emph{i.e.}
\be
    \left[ {\left| B \right\rangle \left\langle C \right|} \right]^{ - 1}  =  - \left| B \right\rangle \left\langle
    C \right|,
\ee
then it is invariant under the interchange $B \leftrightarrow C$:
\be
    \left| B \right\rangle \left\langle C \right| = \left| C \right\rangle \left\langle B
    \right|.
\ee

One can now proceed to develop \textbf{\textit{Sp}} vector
identities in analogy with ordinary vector spaces. For example, it
is useful to note that
\be
    \left| A \right\rangle \left\langle {B}
 \mathrel{\left | {\vphantom {B C}}
 \right. \kern-\nulldelimiterspace}
 {C} \right\rangle  = \left\langle {B}
 \mathrel{\left | {\vphantom {B C}}
 \right. \kern-\nulldelimiterspace}
    {C} \right\rangle \left| A \right\rangle
\ee
leads to the `BAC-CAB' rule:
\be
    \left| A \right\rangle \left\langle {B}
 \mathrel{\left | {\vphantom {B C}}
 \right. \kern-\nulldelimiterspace}
 {C} \right\rangle  = \left\langle {B}
 \mathrel{\left | {\vphantom {B A}}
 \right. \kern-\nulldelimiterspace}
 {A} \right\rangle \left| C \right\rangle  - \left\langle {C}
 \mathrel{\left | {\vphantom {C A}}
 \right. \kern-\nulldelimiterspace}
    {A} \right\rangle \left| B \right\rangle.
\ee

\subsection{The space of complex structure moduli as a special K\"{a}hler manifold}

As promised, we discuss the second definition of special
K\"{a}hler manifolds. Furthermore, since special geometry turns
out to be the same geometry that describes the space $\M_C$ of
complex structure moduli of a CY manifold, we also make the
connection and unify the notation.

The definition goes like this: Let $\mathcal{L}\rightarrow \M$
denote the complex line bundle whose first Chern class equals the
K\"{a}hler form $\K$ of the Hodge-K\"{a}hler manifold $\M$. Now
consider an additional holomorphic flat vector bundle of rank
$(2\htwo+2)$ with structural group $Sp(2\htwo+2, \mathbb{R})$ on
$\M$: $\mathcal{SV}\rightarrow \M$. Construct a tensor bundle
$\mathcal{H}=\mathcal{SV}\otimes \mathcal{L}$. This then is a
special K\"{a}hler manifold if for some holomorphic section
$\left| \Psi  \right\rangle $ of such a bundle (which is a
symplectic vector in the sense of the last section) the K\"{a}hler
2-form is given by:
\be
    K = -\frac{i}{{2\pi }}\partial \bar \partial \ln \left( {i\left\langle {\Psi }
 \mathrel{\left | {\vphantom {\Psi  { \bar\Psi }}}
 \right. \kern-\nulldelimiterspace}
 {{ \bar\Psi }} \right\rangle } \right),
\ee
or in terms of the K\"{a}hler potential on $\M_C$:
\be
    \K =  - \ln \left( {i\left\langle {\Psi }
 \mathrel{\left | {\vphantom {\Psi  { \bar\Psi }}}
 \right. \kern-\nulldelimiterspace}
 {{ \bar\Psi }} \right\rangle } \right)\quad  \to \quad \left\langle {\bar\Psi }
 \mathrel{\left | {\vphantom {\bar\Psi  { \Psi }}}
 \right. \kern-\nulldelimiterspace}
 {{ \Psi }} \right\rangle  = ie^{ - \K}.\label{pot1} \ee

Note that the metric on the bundle is defined via a relation
analogous to (\ref{hodge}). Now, this exactly describes the space
of complex structure moduli $\M_C$ if one chooses:
\be
    \left| \Psi  \right\rangle  = \left( {\begin{array}{*{20}c}
   {Z^I }  \\
   {F_I }  \\
    \end{array}} \right),\quad \quad I = 0, \ldots ,h_{2,1}+1\label{periodvector}
\ee
which, via (\ref{pot1}), leads directly to equation (\ref{pot})
defining the K\"{a}hler potential of $\M_C$. We then identify
$\M_C$ as a special K\"{a}hler manifold with metric $G_{i \bar
j}$. Henceforth, we continue our discussion of $\M_C$ using the
language of special K\"{a}hler geometry and
$Sp(2\htwo+2,\mathbb{R})$ covariance.

Certain constraints on the \textbf{\textit{Sp}} vector $\left|
\Psi \right\rangle$ are imposed as part of the definition, or,
from the point of view of $\M_C$, can also follow as consequences
of equations (\ref{Omegarelations}); these are
\bea
 \left\langle {\Psi }
 \mathrel{\left | {\vphantom {\Psi  {\partial _i \Psi }}}
 \right. \kern-\nulldelimiterspace}
 {{\partial _i \Psi }} \right\rangle  &=& 0 \nonumber\\
 \left\langle {{\nabla _i \Psi }}
 \mathrel{\left | {\vphantom {{\nabla _i \Psi } {\nabla _j \Psi }}}
 \right. \kern-\nulldelimiterspace}
 {{\nabla _j \Psi }} \right\rangle  &=& 0.
\eea

Now, it can be easily demonstrated that the matrix:
\be\label{symplecticmatrix1}
    {\bf \Lambda } = \left[ {\begin{array}{*{20}c}
    {\gamma ^{IK} \theta _{KJ} } & -{\gamma ^{IJ} } \\
    {\left( {\gamma _{IJ}  + \gamma ^{KL} \theta _{IK} \theta _{JL} } \right)} &  - {\gamma ^{JK} \theta _{KI} } \\
    \end{array}} \right]
\ee
satisfies the symplectic condition (\ref{Sp condition 1}), where
$\gamma$ and $\theta$ are defined by (\ref{gammathetadefined}).
Its inverse is then
\be
    {\bf \Lambda }^{ - 1}  =-\bf \Lambda= \left[ {\begin{array}{*{20}c}
    - {\gamma ^{JK} \theta _{KI} } & {\gamma ^{IJ} }  \\
    -{\left( {\gamma _{IJ}  + \gamma ^{KL} \theta _{IK} \theta _{JL} } \right)} &  {\gamma ^{IK} \theta _{KJ} } \\
    \end{array}} \right].
\ee

The symplectic structure manifest here is a consequence of the
topology of the Calabi-Yau manifold $\M$, the origins of which can
be traced to the completeness relations (\ref{cohbasis}), clearly:
\be
   \int\limits_\M {\left[ {\begin{array}{*{20}c}
   {\alpha _I  \wedge \alpha _J } & {\alpha _I  \wedge \beta ^J }  \\
   {\beta ^I  \wedge \alpha _J } & {\beta ^I  \wedge \beta ^J }  \\
    \end{array}} \right]}= \left[ {\begin{array}{*{20}c}
   0 & {\delta _I^J }  \\
   { - \delta _J^I } & 0  \\
    \end{array}} \right] = {\bf S}.
\ee

In fact, if one defines the symplectic vector:
\be
    \left| \Theta  \right\rangle  = \left( {\begin{array}{*{20}c}
   {\beta ^I }  \\
   {\alpha _I }  \\
    \end{array}} \right),
\ee
then it is easy to check that
\be
    \int\limits_\M {{\bf \Theta }\mathop  \otimes \limits_ \wedge  {\bf \Theta }^T }  = {\bf S}^T \quad  \to \quad \int\limits_\M {\left| \Theta  \right\rangle \mathop  \wedge  \left\langle \Theta  \right|}  =  - \mathbbm{1},
\ee
and that the Hodge duality (\ref{formsHodgeduality}) is equivalent
to a rotation in symplectic space:
\be
    \left| {\star\Theta } \right\rangle  = \left| {\Lambda  \Theta }
    \right\rangle.
\ee

Also note that (\ref{defomega}) can similarly be rewritten as
\be
    \Omega  = \left\langle { \Theta}
    \mathrel{\left | {\vphantom {\Theta   }}
    \right. \kern-\nulldelimiterspace}
    {\Psi } \right\rangle.
\ee

Next, we construct a basis in \textbf{\textit{Sp}}. Properly
normalized, the periods vector (\ref{periodvector}) provides such
a basis:
 \be
    \left| V \right\rangle  = e^{\frac{\K}{2}} \left| \Psi  \right\rangle  = \left( {\begin{array}{*{20}c}
   {L^I }  \\
   {M_I }  \\
    \end{array}} \right),
\ee
such that, using (\ref{pot1}):
\be
    \left\langle {{\bar V}}
 \mathrel{\left | {\vphantom {{\bar V} V}}
 \right. \kern-\nulldelimiterspace}
 {V} \right\rangle  = \left( {L^I \bar M_I  - \bar L^I M_I } \right) =
 i.\label{SpBasisNorm}
\ee

From the point of view of physics, equation (\ref{SpBasisNorm}) is the
condition required to obtain an $\N=2$ SUGRA action in the
Einstein frame. If it were not true, then the Einstein-Hilbert
term would have the form
\be
    i\left\langle {{ V}}
    \mathrel{\left | {\vphantom {{ V} \bar V}}
    \right. \kern-\nulldelimiterspace}
    {\bar V} \right\rangle \sqrt {\left| g \right|} R.
\ee

Since ${\left| V \right\rangle }$ is a scalar in the
$\left(i,j,k\right)$ indices, it couples only to the
$U\left(1\right)$ bundle via the K\"{a}hler covariant derivative
(\ref{U1 cov der}) as follows:
\bea
     \left|\nabla _i V \right\rangle  &=& \left|\left[ {\partial _i  + \frac{1}{2}\left( {\partial _i \K} \right)} \right] V \right\rangle ,\quad \quad  \left|\nabla _{\bar i} V \right\rangle  =\left| \left[ {\partial _{\bar i}  - \frac{1}{2}\left( {\partial _{\bar i} \K} \right)} \right] V \right\rangle  \nonumber\\
     \left|\nabla _i {\bar V} \right\rangle  &=& \left|\left[ {\partial _i  - \frac{1}{2}\left( {\partial _i \K} \right)} \right] {\bar V} \right\rangle ,\quad \quad  \left|\nabla _{\bar i} {\bar V} \right\rangle  = \left|\left[ {\partial _{\bar i}  + \frac{1}{2}\left( {\partial _{\bar i} \K} \right)} \right] {\bar V}
    \right\rangle.
\eea

In other words, the K\"{a}hler weights of ${\left| V \right\rangle
}$ are $\left(1,-1\right)$. Using this, one can construct the
orthogonal \textbf{\textit{Sp}} vectors:
\bea
    \left| {U_i } \right\rangle  &=& \left| \nabla _i V
    \right\rangle  =  \left(
    {\begin{array}{*{20}c}
   {\nabla _i L^I }  \\
   {\nabla _i M_I }  \\
    \end{array}} \right) = \left( {\begin{array}{*{20}c}
   {f_i^I }  \\
   {h_{i|I} }  \\
    \end{array}} \right) \\
     \left| {U_{\bar i} } \right\rangle  &=& \left|\nabla _{\bar i}  {\bar V} \right\rangle  =\left( {\begin{array}{*{20}c}
   {\nabla _{\bar i} \bar L^I }  \\
   {\nabla _{\bar i} \bar M_I }  \\
    \end{array}} \right) = \left( {\begin{array}{*{20}c}
   {f_{\bar i}^I }  \\
   {h_{\bar i|I} }  \\
    \end{array}} \right),
\eea
with the same K\"{a}hler weights as ${\left| V \right\rangle }$,
\emph{i.e.}
\bea
     \left|\nabla _i U_j  \right\rangle  &=& \left|\left[ {\partial _i  + \frac{1}{2}\left( {\partial _i \K} \right)} \right] U_j  \right\rangle ,\quad \quad  \left|\nabla _{\bar i} U_j  \right\rangle  =\left| \left[ {\partial _{\bar i}  - \frac{1}{2}\left( {\partial _{\bar i} \K} \right)} \right] U_j  \right\rangle  \nonumber\\
     \left|\nabla _i U_{\bar j} \right\rangle  &=& \left|\left[ {\partial _i  - \frac{1}{2}\left( {\partial _i \K} \right)} \right] U_{\bar j} \right\rangle ,\quad \quad  \left|\nabla _{\bar i} U_{\bar j} \right\rangle  = \left|\left[ {\partial _{\bar i}  + \frac{1}{2}\left( {\partial _{\bar i} \K} \right)} \right] U_{\bar j}
    \right\rangle.
\eea

Note that $\left| {U_i } \right\rangle$ also couples to the metric
$G_{i\bar j}$ via the Levi-Civita connection. So its full
covariant derivative is defined by (\ref{CovariantLeviKahler}):
\bea
    \left| {\D_i U_j } \right\rangle  &=& \left| {\nabla _i U_j } \right\rangle  - \Gamma _{ij}^k \left| {U_k } \right\rangle \quad \quad \left| {\D_{\bar i} U_j } \right\rangle  = \left| {\nabla _{\bar i} U_j } \right\rangle  \nonumber\\
    \left| {\D_i U_{\bar j} } \right\rangle  &=& \left| {\nabla _i U_{\bar j} } \right\rangle \quad \quad \quad\quad\quad\quad\;\left| {\D_{\bar i} U_{\bar j} } \right\rangle  = \left| {\nabla _{\bar i} U_{\bar j} } \right\rangle  - \Gamma _{\bar i\bar j}^{\bar k} \left| {U_{\bar k} }
    \right\rangle.
\eea

It can be demonstrated that these quantities satisfy the
properties
\bea
    \left|\nabla _i  {\bar V} \right\rangle  &=& \left|\nabla _{\bar i}  V \right\rangle =0\label{Normality2}\\
    \left\langle {{U_i }}
    \mathrel{\left | {\vphantom {{U_i } {U_j }}}
    \right. \kern-\nulldelimiterspace}
    {{U_j }} \right\rangle  &=& \left\langle {{U_{\bar i} }}
    \mathrel{\left | {\vphantom {{U_{\bar i} } {U_{\bar j} }}}
    \right. \kern-\nulldelimiterspace}
    {{U_{\bar j} }} \right\rangle    =0\\
    \left\langle {\bar V}
    \mathrel{\left | {\vphantom {\bar V {U_i }}}
    \right. \kern-\nulldelimiterspace}
    {{U_i }} \right\rangle  &=& \left\langle {V}
    \mathrel{\left | {\vphantom {V {U_{\bar i} }}}
    \right. \kern-\nulldelimiterspace}
    {{U_{\bar i} }} \right\rangle  = \left\langle { V}
    \mathrel{\left | {\vphantom { V {U_i }}}
    \right. \kern-\nulldelimiterspace}
    {{U_i }} \right\rangle=\left\langle {\bar V}
    \mathrel{\left | {\vphantom {\bar V {U_{\bar i} }}}
    \right. \kern-\nulldelimiterspace}
    {{U_{\bar i} }} \right\rangle= 0,\label{Normality}\\
    \left|\nabla _{\bar j}  {U_i } \right\rangle  &=& G_{i\bar j} \left| V \right\rangle ,\quad \quad \left|\nabla _i  {U_{\bar j} } \right\rangle  = G_{i\bar j} \left| {\bar V}
    \right\rangle,\\
    G_{i\bar j}&=& \left( {\partial _i \partial _{\bar j} \K} \right)=- i    \left\langle {{U_i }}
    \mathrel{\left | {\vphantom {{U_i } {U_{\bar j} }}}
    \right. \kern-\nulldelimiterspace}
    {{U_{\bar j} }} \right\rangle.\label{KmetricasSpproduct}
\eea

Note that (\ref{Normality2}) implies
\be
    \left|\partial _i  {\bar \Psi } \right\rangle  =\left|\partial _{\bar i}  \Psi  \right\rangle  =
    0.
\ee

This definition of special K\"{a}hler manifolds is directly
related to the first definition in \S\ref{SKM First} via the
identification:
\be
    \left| {\D _i U_j } \right\rangle  = G^{k\bar l} C_{ijk} \left| {U_{\bar l} }
    \right\rangle,
\ee
which leads to:
\be
    C_{ijk}  = -i\left\langle {{\D _i U_j }}
    \mathrel{\left | {\vphantom {{\D _i U_j } {U_k }}}
    \right. \kern-\nulldelimiterspace}
    {{U_k }} \right\rangle.
\ee

The following identities may now be derived:
\bea
    \mathcal{N}_{IJ} L^J  &=& M_I ,\quad \quad \quad \quad \mathcal{\bar N}_{IJ} f_i^J  = h_{i|I}  \nonumber\\
    \mathcal{\bar N}_{IJ} {\bar L}^J  &=& {\bar M}_I ,\quad \quad \quad \quad \mathcal{ N}_{IJ}  f_{\bar i}^J  =  h_{{\bar i}|I}  \label{PeriodMatrix}\\
    \gamma _{IJ} L^I \bar L^J  &=& \frac{1}{2},\quad \quad \quad \quad G_{i\bar j}  = 2\gamma _{IJ} f_i^I f_{\bar j}^J,  \label{gammametric}
\eea
as well as the very useful
\bea
    \gamma ^{IJ}  &=& 2\left( L^I \bar L^J + {G^{i\bar j} f_i^I f_{\bar j}^J   } \right)  \nonumber\\
    \left( {\gamma _{IJ}  + \gamma ^{KL} \theta _{IK} \theta _{JL} } \right) &=& 2\left( { M_I \bar M_J + G^{i\bar j} h_{i|I} h_{\bar j|J} } \right) \nonumber\\
    \gamma ^{IK} \theta _{KJ}&=&2\left(\bar L^I M_J + G^{i\bar j}f_i^{\,\,I}h_{\bar
    j|J}\right)+i \delta^I_J\nonumber\\
    &=&2\left(L^I \bar M_J + G^{i\bar j}h_{i|J} f_{\bar j}^{\,\,\,I}\right)-i\delta^I_J\nonumber\\
    &=& \left({L^I \bar M_J  + \bar L^I M_J } \right) + G^{i\bar j} \left( {f_i^{\,\,I}h_{\bar j|J}  + h_{i|J} f_{\bar j}^{\,\,\,I} } \right).\label{GammaThetaLMconnection}
\eea

Note that the last formula in (\ref{gammametric}) in particular
implies that the imaginary part of the period matrix
$\left({\mathop{\rm Im}\nolimits} \mathcal{N}_{IJ} =  -
\gamma_{IJ}\right)$ acts as a metric in the $\left(I,J,K\right)$
indices, and that $f_i^I$ are the vielbeins relating it to the
special K\"{a}hler metric $G_{i\bar j}$ similar to (\ref{bein}).
In other words, these relations provide a connection between the
SKG structure and the \textbf{\textit{Sp}} space. We will now
exploit this. Equations (\ref{GammaThetaLMconnection}) lead to a
second form for the symplectic matrix (\ref{symplecticmatrix1}):
\be\label{symplecticmatrix2}
    {\bf \Lambda } = \left[ {\begin{array}{*{20}c}
        {\left({L^I \bar M_J  + \bar L^I M_J } \right)} & {} & {-{2\left( {L^I \bar L^J+G^{i\bar j} f_i^I f_{\bar j}^J   } \right)}}  \\
        {+{G^{i\bar j} \left( {f_i^{\,\,I}h_{\bar j|J}  + h_{i|J} f_{\bar j}^{\,\,\,I} } \right)}} & {}  \\
        {} & {} & {-\left({L^J \bar M_I  + \bar L^J M_I } \right)}  \\
        {2\left( {M_I \bar M_J+G^{i\bar j} h_{i|I} h_{\bar j|J}   } \right)} & {} & {-G^{i\bar j} \left( {f_i^{\,\,J}h_{\bar j|I}  + h_{i|I} f_{\bar j}^{\,\,\,J} } \right)}  \\
    \end{array}} \right]
\ee
with inverse
\be
    {\bf \Lambda }^{ - 1}  = -{\bf \Lambda } =\left[ {\begin{array}{*{20}c}
        {-\left({L^J \bar M_I  + \bar L^J M_I } \right)} & {} & {{2\left( {L^I \bar L^J+G^{i\bar j} f_i^I f_{\bar j}^J   } \right)}}  \\
        {-G^{i\bar j} \left( {f_i^{\,\,J}h_{\bar j|I}  + h_{i|I} f_{\bar j}^{\,\,\,J} } \right)} & {} & {}  \\
        {} & {} & {\left({L^I \bar M_J  + \bar L^I M_J } \right)}  \\
        {-2\left( {M_I \bar M_J +G^{i\bar j} h_{i|I} h_{\bar j|J}  } \right)} & {} & {+{G^{i\bar j} \left( {f_i^{\,\,I}h_{\bar j|J}  + h_{i|J} f_{\bar j}^{\,\,\,I} } \right)}}  \\
    \end{array}} \right].
\ee

By inspection, one can write down the following important result:
\bea
    {\bf \Lambda } &=& \left| V \right\rangle \left\langle {\bar V} \right| + \left| {\bar V} \right\rangle \left\langle V \right| + G^{i\bar j} \left| {U_i } \right\rangle \left\langle {U_{\bar j} } \right| + G^{i\bar j} \left| {U_{\bar j} } \right\rangle \left\langle {U_i } \right| \nonumber\\
    {\bf \Lambda }^{ - 1}  &=&  - \left| V \right\rangle \left\langle {\bar V} \right| - \left| {\bar V} \right\rangle \left\langle V \right| - G^{i\bar j} \left| {U_i } \right\rangle \left\langle {U_{\bar j} } \right| - G^{i\bar j} \left| {U_{\bar j} } \right\rangle \left\langle {U_i } \right|.\label{Lambdaasouter1}
\eea

In other words, the rotation matrix in \textbf{\textit{Sp}} is
expressible as the outer product of the basis vectors. Note that since $\bf \Lambda$
satisfies the property (\ref{LambdaProperty}), it is invariant
under the interchange $V \leftrightarrow \bar V$ and/or $U_i
\leftrightarrow U_{\bar j}$. This makes manifest the fact that
$\bf \Lambda$ is a real matrix; ${\bf \Lambda } = {\bf \bar
\Lambda }$. Now, applying ${\bf \Lambda }^{ - 1} {\bf \Lambda } =
\mathbbm{1}$, we end up with the condition
\be
    \left| {\bar V} \right\rangle \left\langle V \right| + G^{i\bar j} \left| {U_i } \right\rangle \left\langle {U_{\bar j} } \right|=\left| V \right\rangle \left\langle {\bar V} \right|+G^{i\bar j} \left| {U_{\bar j} } \right\rangle \left\langle {U_i } \right|-i,
\ee
which can be checked explicitly using
(\ref{GammaThetaLMconnection}). This can be used to write $\bf
\Lambda$ in an even simpler form:
\bea
    {\bf \Lambda } &=& 2\left| V \right\rangle \left\langle {\bar V} \right| + 2G^{i\bar j} \left| {U_{\bar j} } \right\rangle \left\langle {U_i } \right|
    -i\nonumber\\
    {\bf \Lambda }^{-1} &=& -2\left| V \right\rangle \left\langle {\bar V} \right| - 2G^{i\bar j} \left| {U_{\bar j} } \right\rangle \left\langle {U_i } \right|
    +i.\label{Lambdaasouter2}
\eea

It can further be shown that
\bea
    \D_i {\bf \Lambda } =\nabla_i {\bf \Lambda }= \partial_i {\bf \Lambda } = 2\left| {U_i } \right\rangle \left\langle {\bar V} \right|+2\left| {\bar V} \right\rangle \left\langle {U_i } \right| + 2G^{j\bar r} G^{k\bar p} C_{ijk} \left| {U_{\bar r} } \right\rangle \left\langle {U_{\bar p} } \right|.
    \label{CovariantDerofLambda}
\eea

Finally, we note that our discussion here is based on a
definition of special manifolds that is not the only one in
existence. See, for instance, \cite{Craps:1997gp} for details.
Explicit examples of special manifolds in various dimensions are
given in, for example, \cite{hep-th/9512043}.

\section{$D=5$ $\N=2$ supergravity with hypermultiplets\label{dimensionalreduction}}

In this section, we review the derivation of ungauged $D=5$ $\N=2$
SUGRA via the dimensional reduction of $D=11$ SUGRA over a
Calabi-Yau manifold $\M$. Specifically, we look at the case where
only the complex structure of $\M$ is deformed. For
the sake of compactness and clarity, we emphasize the use of
differential forms on the spacetime manifold, following the
definitions in the appendix.

\subsection{Dimensional reduction}

The unique supersymmetric gravity theory in eleven dimensions has
the following bosonic action:
\be
    S_{11}  = \int_{11} \left( { {\mathcal{R}\star 1 -
    \frac{1}{2}\F \wedge \star \F -
    \frac{1}{6}\A \wedge \F \wedge
    \F}}\right),
\ee
where $\mathcal{R}$ is the $D=11$ Ricci scalar, $\A$ is the 3-form
gauge potential and $\F=d\A$. The dimensional reduction is
traditionally done using the metric:
\bea
    ds^2 &=& G_{MN} dx^M dx^N \nonumber \\
    &=& e^{\frac{2}{3}\sigma } g_{\mu \nu } dx^\mu  dx^\nu
    + e^{ - \frac{\sigma }{3}} k_{ab} dx^a dx^b \nonumber\\
    & & M,N = 0, \ldots, 10 \quad \mu ,\nu  = 0, \ldots ,4 \quad \quad a,b = 1, \ldots
    ,6,\label{expmet}
\eea
where $g_{\mu \nu }$ is the target five dimensional metric,
$k_{ab}$ is a metric on the six dimensional compact subspace $\M$,
the dilaton $\sigma$ is a function in $x^\mu$ only and the warp
factors are chosen to give the conventional coefficients
in five dimensions, guaranteeing that the gravitational term in the action will have the standard Einstein-Hilbert form. We choose a complex structure on $\M$ such
that
\be
    k_{ab} dx^a dx^b=k_{mn} dw^m dw^n  + k_{\bar m\bar n} dw^{\bar m} dw^{\bar n}  + 2k_{m\bar n} dw^m dw^{\bar
    n},
\ee
where the holomorphic and antiholomorphic indices $(m,n;\bar m
,\bar n )$ are three dimensional on $\M$. The Hermiticity
condition (\ref{Hermitian}) demands that $k_{mn}=k_{\bar m\bar
n}=0$, while the Ricci tensor for $\M$ is set to zero as
dictated by Yau's theorem. Furthermore, we consider the case where
only the complex structure is deformed, which requires $\delta
k_{m\bar n} = 0$ and $\left( {\delta k_{mn} ,\delta k_{\bar m\bar
n} } \right) \ne 0$, as discussed earlier.

Now, the flux compactification of the gauge field is done by
expanding $\A$ into two forms, one is the five dimensional gauge
field $A$ while the other contains the components of $\A$ on $\M$
written in terms of the cohomology forms $\left(
\alpha_I,\beta^I\right)$ as follows:
\bea
    \A &=& A + \sqrt 2 \left( {\zeta ^I \alpha _I  + \tilde \zeta _I \beta ^I }
    \right),\quad\quad \left({I = 0, \ldots ,h_{2,1} } \right)\nonumber\\
    \F &=& d\A = F + F_\mu dx^\mu,\nonumber\\
    F_\mu  dx^\mu   &=& \sqrt 2 \left[ {\left(d\zeta ^I\right)\wedge \alpha _I  + \left(d\tilde \zeta _I\right)\wedge \beta ^I }
    \right]\nonumber\\
    &=&\sqrt 2 \left[ {\left( {\partial _\mu \zeta ^I } \right)\alpha _I  + \left( {\partial _\mu \tilde \zeta _I } \right)\beta ^I } \right]\wedge dx^\mu. \label{FExpanded}
\eea

Because of the eleven dimensional Chern-Simons term, the
coefficients $\zeta^I$ and $\tilde \zeta_I$ appear as
pseudo-scalar axion fields in the lower dimensional theory. We
also note that $A$ in five dimensions is dual to a scalar field
which we will call $a$ (known as the universal axion). The set
($a$, $\sigma$, $\zeta^0$, $\tilde \zeta_0$) is known as the
universal hypermultiplet\footnote{So called because it appears in
all Calabi-Yau compactifications, irrespective of the detailed
structure of the CY manifold. The dilaton $\sigma$
is proportional to the natural logarithm of the volume of $\M$.}.
The rest of the hypermultiplets are ($z^i$, $z^ {\bar i}$,
$\zeta^i$, $\tilde \zeta_i$ : $i = 1, \ldots ,h_{2,1} $), where we
will recognize the $z$'s as the CY's complex structure moduli.
Note that the total number of scalar fields in the hypermultiplets
sector is $4(h_{2,1}+1)$ (each hypermultiplet has 4 real scalar
fields) which comprises a quaternionic manifold as noted earlier.
Also included in the hypermultiplets are the fermionic partners of
the hypermultiplet scalars known as the hyperini (singular:
hyperino). However, in what follows, we will only discuss the
bosonic part of the action. The hyperini, as well as the
gravitini, will make their appearance in the SUSY variation
equations later. The bits and pieces one needs for the dimensional
reduction are as follows:
\begin{enumerate}
  \item The metric components
        \begin{eqnarray}
            G_{\mu \nu }  &=& e^{\frac{2}{3}\sigma } g_{\mu \nu
            },\quad G^{\mu \nu }  = e^{-\frac{2}{3}\sigma } g^{\mu \nu } \nonumber\\
            G_{ab}  &=& e^{ - \frac{\sigma }{3}} k_{ab}, \quad G^{ab}  = e^{ \frac{\sigma }{3}} k^{ab}
            \nonumber \\
            G &=& \det  {G_{MN} }  = e^{\frac{4}{3}\sigma } g k, \nonumber \\
            g &=& \det  {g_{\mu \nu } }  , \quad k =  \det k_{ab}.
        \end{eqnarray}
  \item The Christoffel symbols
        \begin{eqnarray}
            \Gamma _{\nu \rho }^\mu   &=& \tilde \Gamma _{\nu \rho }^\mu  \left[ {g} \right] + \frac{1}{3}\left[ { \delta _\nu ^\mu  \left( {\partial _\rho  \sigma
            }     \right) +\delta _\rho ^\mu  \left( {\partial   _\nu      \sigma } \right)- \delta_{\nu \rho } \delta^{\mu \kappa} \left( {\partial
            _\kappa      \sigma } \right)} \right]
            \nonumber \\
            \Gamma _{ab}^\mu   &=& \frac{1}{6}e^{ - \sigma } k_{ab}
            \left( {\partial            ^\mu \sigma } \right) - \frac{1}{2}e^{ - \sigma }
            \left(     {\partial ^\mu k_{ab}     } \right) \nonumber \\
            \Gamma _{\mu b}^a  &=& \frac{1}{2}k^{ac} \left( {\partial _\mu  k_{cb} } \right)
            -     \frac{1}{6}\delta _b^a \left( {\partial _\mu  \sigma } \right)
            \nonumber     \\
            \Gamma _{bc}^a  &=& \hat \Gamma _{bc}^a \left[ {k} \right],
        \end{eqnarray}
        where the ( $\tilde{}$ ) and the ( $\hat{}$ ) refer to the purely five
        and six dimensional components respectively.
\end{enumerate}

Now, calculating the eleven dimensional Ricci scalar based on this
gives:
\be
    \sqrt {\left|G \right|} \mathcal{R} = \sqrt { \left| gk\right|} \left[ {  R\left[  {g } \right] - \frac{1}{2}\left( {\partial_\mu \sigma } \right)\left( {\partial^\mu \sigma } \right)
    - \frac{1}{4}k^{m\bar n} k^{r\bar p} \left( {\partial _\mu  k_{mr} } \right)\left( {\partial _\mu  k_{\bar n\bar p} } \right) }
    \right],
\ee
where we have used $\hat R_{ab}  = 0$ since $\M$ is Ricci-flat, as
well as dropped all total derivatives and terms containing
${k_{mn} }$, ${k_{\bar m\bar n} }$ and $\delta k_{m\bar n} $.
Using (\ref{Gij}) one gets:
\begin{equation}
    \int_{11} {\mathcal{R}\star 1}  =  \int {d^5 x\sqrt {\left| g \right|}
    \left[     { R  - \frac{1}{2}\left( {\partial _\mu  \sigma } \right)\left(
    {\partial     ^\mu  \sigma } \right) - G_{i\bar j} ( {\partial _\mu  z^i }
    )( {\partial ^\mu  z^{\bar j} } )} \right]},
\end{equation}
where we have normalized the volume of the compact space to
$V_{CY}=1$. Next, the Maxwell term is:
\be
     - \frac{1}{2}\F \wedge  \star \F \rightarrow  - \frac{1}{2}\frac{1}{{4!}}\F_{LMNP} \F^{LMNP}  =  - \frac{1}{{48}}\left( {e^{ - 2\sigma } F_{\mu \nu \rho \sigma } F^{\mu \nu \rho \sigma }  + e^{\frac{2}{3}\sigma } F_\mu  F^\mu  }
     \right).
\ee

Substituting, we get:
\begin{eqnarray}
    -    \frac{1}{2}\int_{11}{\F \wedge \star \F} = &-& \frac{1}{{48}}\int {d^5 x\sqrt { \left| g\right| } e^{
    -     2\sigma } F_{\mu \nu \rho \sigma } F^{\mu \nu \rho \sigma } }  \\
    &-&\int {d^5 x\sqrt { \left| g\right| } e^\sigma  \left[ {\left( {\partial _\mu  \zeta ^I } \right)\left( {\partial
    ^\mu      \zeta ^J } \right)\int\limits_{\M} {\alpha _I  \wedge \star\alpha _J } } \right. }\nonumber \\
    &+&  ( {\partial   _\mu  \zeta ^I } )( {\partial ^\mu  \tilde \zeta _J }
    )\int\limits_{\M} {\alpha _I  \wedge \star\beta ^J }+( {\partial _\mu  \tilde \zeta _I } )( {\partial ^\mu
    \zeta     ^J } )\int\limits_{\M} {\beta ^I  \wedge \star\alpha _J }  \nonumber \\
    &+& \left. { ( {\partial _\mu \tilde \zeta _I } )( {\partial ^\mu  \tilde \zeta _J } )\int\limits_{\M} {\beta
    ^I      \wedge \star\beta ^J } } \right],
\end{eqnarray}
where the Hodge star on the right hand side is with respect to
$\M$. Now, using (\ref{cohbasis}) and (\ref{formsHodgeduality}) we
end up with:
\bea
    -    \frac{1}{2}\int_{11}{\F \wedge \star \F}=
    &-&    \int{d^5 x\sqrt { \left| g\right| }} \left\{ {\frac{1}{{48}}e^{-2\sigma }F_{\mu \nu \rho \sigma }F^{\mu \nu \rho \sigma}} \right.
     \nonumber \\
    &+& e^\sigma  \left[ {\left( {\gamma_{IJ}  + \gamma ^{KL} \theta_{IK}\theta_{JL} } \right)\left(
    {\partial _\mu  \zeta^I } \right)\left( {\partial ^\mu  \zeta^J } \right) + \gamma ^{ IJ}     ( {\partial _\mu  \tilde \zeta_I } )( {\partial ^\mu  \tilde
    \zeta_J     } )} \right. \nonumber \\
    &+& \left. {\left. {2\gamma ^{ IK} \theta_{JK} ( {\partial _\mu  \zeta^J } )(
    {\partial ^\mu  \tilde \zeta_I } )} \right]} \right\}.
\eea

Finally, the Chern-Simons term gives:
\bea
  - \frac{1}{6}\int_{11}{\A \wedge \F \wedge \F}  &=&  - \int_5 {\left[ {\zeta ^I F \wedge d\tilde \zeta _J   \int\limits_\M {\alpha _I  \wedge \beta ^J }  - \tilde \zeta _I F \wedge d\zeta ^J   \int\limits_\M {\alpha _J  \wedge \beta ^I } } \right]}  \nonumber\\
  &=&  - \int_5 { F \wedge \left( {\zeta ^I d\tilde \zeta _I  - \tilde \zeta _I d\zeta ^I }
  \right)}.
\eea

To sum up, the ungauged five dimensional $\N=2$ supergravity
bosonic action with vanishing vector multiplets is:
\bea
  S_5  &=& \int\limits_5 \left\{ {{R\star 1 - \frac{1}{2}d\sigma \wedge\star d\sigma  - G_{i\bar j} dz^i \wedge\star dz^{\bar j}  - F \wedge \left( {\zeta^I d\tilde \zeta_I  - \tilde \zeta_I d\zeta^I } \right)} - \frac{1}{2}e^{ - 2\sigma } F \wedge \star F} \right. \nonumber\\
  &-&  \left. {e^\sigma  \left[ {\left( {\gamma_{IJ}  + \gamma ^{ KL} \theta _{IK}\theta _{JL} } \right) {d\zeta^I } \wedge\star {d\zeta^J }  + \gamma ^{IJ}  {d\tilde \zeta_I } \wedge\star {d\tilde \zeta_J }  + 2\gamma ^{ IK} \theta_{JK}  {d\zeta^J } \wedge\star {d\tilde \zeta_I } }
  \right]} \right\}. \label{Fullaction}
\eea

To complete the picture, we vary the action and present the field
equations of $\sigma$, $\left(z^i,z^{\bar i}\right)$, $A$ and
$\left(\zeta^I,\tilde\zeta_I\right)$ respectively:
\bea
    \left( {\Delta \sigma } \right)\star 1 - e^\sigma  X + e^{ - 2\sigma } F \wedge \star F &=& 0\label{dilatoneom}\\
    \left( {\Delta z^i } \right)\star 1 + \Gamma _{jk}^i dz^j  \wedge \star dz^k  - \frac{1}{2}e^\sigma  G^{i\bar j} \left( {\partial _{\bar j} X} \right)\star 1 &=& 0 \nonumber\\
    \left( {\Delta z^{\bar i} } \right)\star 1 + \Gamma _{\bar j\bar k}^{\bar i} dz^{\bar j}  \wedge \star dz^{\bar k}  - \frac{1}{2}e^\sigma  G^{\bar ij} \left( {\partial _j X} \right)\star 1 &=& 0\label{zeom} \\
    d^{\dag} \left[ {e^{ - 2\sigma } F + \star\left( {\zeta ^I d\tilde \zeta _I  - \tilde \zeta _I d\zeta ^I } \right)} \right] &=& 0\label{Feomgeneral}\\
    d^\dag\left[ e^\sigma  \gamma ^{ IK} \theta_{JK}   {d  \zeta^J }  + e^\sigma  \gamma ^{ IJ}
    {d  \tilde \zeta_J }+ \zeta^I \star  F\right]&=&0\nonumber \\
    d^\dag\left[ e^\sigma  \left( {\gamma_{IJ}  + \gamma ^{ KL} \theta _{IK}\theta _{JL} } \right) {d  \zeta^J }  + e^\sigma \gamma ^{ JK} \theta_{IK}  {d  \tilde \zeta_J }   - \tilde \zeta_I \star  F
    \right]&=&0,\label{xieom}
\eea
where for compactness we have defined
\be
    X= {\left( {\gamma_{IJ}  + \gamma ^{ KL} \theta _{IK}\theta _{JL} } \right) {d\zeta^I } \wedge\star {d\zeta^J }  + \gamma ^{IJ}  {d\tilde \zeta_I } \wedge\star {d\tilde \zeta_J }  + 2\gamma ^{ IK} \theta_{JK}  {d\zeta^J } \wedge\star {d\tilde \zeta_I } }
  ,\label{X}
\ee
as well as used the Bianchi identity $d F=0$ to get the given
form of (\ref{xieom}). From a five dimensional perspective, the
moduli $\left(z^i,z^{\bar i}\right)$ behave as scalar fields. We
recall, however, that the behavior of the other fields is
dependent on the moduli, \emph{i.e.} they are functions in them.
Hence it is possible to treat (\ref{zeom}) as constraints that can
be used to reduce the degrees of freedom of the other field
equations.

Equations (\ref{Feomgeneral}) and (\ref{xieom}) are clearly the
statements that the forms:
\bea
    \mathcal{J}_2  &=& e^{ - 2\sigma } F + \star\left( {\zeta ^I d\tilde \zeta _I  - \tilde \zeta _I d\zeta ^I } \right)\nonumber\\
    \mathcal{J}_5^I  &=&  e^\sigma  \gamma ^{ IK} \theta_{JK}   {d  \zeta^J }  + e^\sigma  \gamma ^{ IJ}
    {d  \tilde \zeta_J }+ \zeta^I\star F\nonumber\\
    \mathcal{\tilde J}_{5|I}  &=& e^\sigma  \left( {\gamma_{IJ}  + \gamma ^{ KL} \theta _{IK}\theta _{JL} } \right) {d  \zeta^J }  + e^\sigma \gamma ^{ JK} \theta_{IK}  {d  \tilde \zeta_J }   -\tilde \zeta_I \star  F\label{Currents}
\eea
are conserved. These are, in fact, Noether currents corresponding
to certain isometries of the quaternionic manifold defined by the
hypermultiplets as discussed in various sources
\cite{Ferrara:1989ik, Cecotti:1988qn}. From a five dimensional
perspective, they can be thought of as the result of the
invariance of the action under particular infinitesimal shifts of
$A$ and $\left(\zeta, \tilde\zeta\right)$
\cite{Gutperle:2000ve,Gutperle:2000sb}. The charge densities
corresponding to them can then be found in the usual way by:
\be
 \mathcal{Q}_2  = \int {\mathcal{J}_2 },\quad \quad \quad
 \mathcal{Q}_5^I  = \int { \mathcal{J}_5^I} ,\quad \quad \quad  \mathcal{\tilde Q}_{5|I}  = \int { \mathcal{\tilde J}_{5|I} }.\label{Charges}
\ee

The geometric way of understanding these charges is noting that
they descend from the eleven dimensional electric and magnetic
M-brane charges, hence the $\left(2,5\right)$ labels\footnote{This
is the reverse situation to that of \cite{Gutperle:2000ve}, where
the (dual) Euclidean theory was studied.}. M2-branes wrapping
special Lagrangian cycles of $\M$ generate $\mathcal{Q}_2$ while
the wrapping of M5-branes excite
$\left(\mathcal{Q}_5^I,\mathcal{\tilde Q}_{5|I}\right)$.

Finally, for completeness sake we also give $da$, where $a$ is the
universal axion dual to $A$. Since (\ref{Feomgeneral}) is
equivalent to $d^2 a =0$, we conclude that
\be\label{UniversalAxion}
    da = e^{ - 2\sigma } \star F - \left( {\zeta ^I d\tilde \zeta _I  - \tilde \zeta _I d\zeta ^I } \right),
\ee
where $a$ is governed by the field equation
\be\label{a field equation}
    d^{\dag} \left[ {e^{2\sigma } da + e^{2\sigma } \left( {\zeta ^I d\tilde \zeta _I  - \tilde \zeta _I d\zeta ^I } \right)} \right] =    0;
\ee
as a consequence of $dF=0$. Both terms involving $F$ in
(\ref{Fullaction}) could then be replaced by the single
expression\footnote{Alternatively, one may dualize the action by
introducing $a$ as a Lagrange multiplier and modifying the action
accordingly \cite{Gutperle:2000sb}.}
\be
    S_a  = \frac{1}{2}\int {e^{2\sigma } \left[ {da + \left( {\zeta^I d\tilde \zeta_I  - \tilde \zeta_I d\zeta^I } \right)} \right] \wedge \star\left[ {da + \left( {\zeta^I d\tilde \zeta_I  - \tilde \zeta_I d\zeta^I } \right)}
    \right]}.\label{a action}
\ee

\subsection{Supersymmetry}

In this section we briefly outline the derivation of the five
dimensional SUSY variation equations from the eleven dimensional
one. As before, enough material is reviewed for an overall
understanding rather than a detailed description. On a CY 3-fold,
there are two supercovariantly constant Killing spinors \cite{De
Wit:1985qn}, that may be defined, as usual, in terms of the Dirac
matrices acting as `creation' and `annihilation' operators on the
spinors as follows:
\begin{equation}\label{}
    \Gamma _{\hat m} \eta_+  = 0 ,\quad \Gamma _{\hat {\bar m}} \eta_-  =
    0,
\end{equation}
where once again the hatted indices are flat indices on a tangent
space. It follows then that
\begin{equation}\label{projections}
    \Gamma _{\hat m\hat n\hat p} \eta_+  =  \pm  \bar\varepsilon
    _{\hat m\hat n\hat p} \eta_-,
\end{equation}
where $\Gamma _{\hat m\hat n\hat p}$ is the antisymmetrized
product of $\Gamma _{\hat m}$. We use (\ref{projections}) to
define the spinors in terms of the CY $(3,0)$ form as follows:
\begin{eqnarray}
    \Gamma _{mnp} \eta_+  &=& \Omega _{mnp} \eta_-  \nonumber\\
    \eta_-  &=& \frac{1}{{3!\left\| \Omega  \right\|^2 }}{\bar\Omega} ^{mnp}
    \Gamma _{mnp} \eta_+.
\end{eqnarray}

Now, the eleven dimensional $\N=1$ spinor $\Pi $ may be expanded
in terms of the five dimensional $\N=2$ spinors $\epsilon_1$ and
$\epsilon_2$ as follows:
\begin{equation}\label{spinorexpansion}
    \Pi  = \epsilon _1  \otimes \eta_+  + \epsilon _2 \otimes
     \eta_-.
\end{equation}

The strategy is to write the eleven dimensional gravitino
equation, expand in terms of the five dimensional spinors
similarly to (\ref{spinorexpansion}), then identify the terms that
are dependent on the $(2,1)$ and $(1,2)$ forms $\chi_i$ and
$\chi_{\bar i}$, or, via Kodaira's formula (\ref{Kodaira}),
$\left( {\nabla _i \Omega } \right)$ and $\left( {\nabla _{\bar i}
\bar \Omega } \right)$. These are taken to represent the hyperini,
and their sum is identified as the hyperino variation equations.
The rest of the terms, dependent on the $(3,0)$ and $(0,3)$ forms,
become the $\mathcal{N}=2$ gravitini equations.

We begin with the $D=11$ gravitino variation:
\begin{equation}
    \delta_{\Pi} \psi _M = \left( {\partial _M \Pi } \right) + \frac{1}{4}\omega _M^{\,\,\,\,\,\,\hat L\hat N} \Gamma _{\hat L\hat N}\Pi -
    \frac{1}{{288}}F_{LNPQ } \left( {\Gamma _M  ^{\;\;\;\;LNPQ }  - 8\delta _M
    ^L  \Gamma ^{NPQ } } \right)\Pi.\label{eleventranslambda}
\end{equation}

Based on the metric (\ref{expmet}), we collect the relevant beins:
\begin{eqnarray}
    e_{\;\;\nu} ^{\hat \mu }  = e^{\frac{\sigma}{3} } N_{\;\;\nu} ^{\hat \mu }   &,&
    \quad e_{\;\;b}^{\hat a}  = e^{ - \frac{\sigma}{6} } W_{\;\;b}^{\hat a}  \nonumber \\
    g_{\mu \nu }  = N_{\;\;\mu} ^{\hat \alpha } N_{\;\;\nu} ^{\hat \beta } \eta _{\hat \alpha \hat
    \beta     }   &,&\quad  k_{ab}  = W_{\;\;a}^{\hat c} W_{\;\;b}^{\hat d} \delta _{\hat c\hat
    d}.
\end{eqnarray}
The non-vanishing components of the spin connections are then:
\begin{eqnarray}
    \omega _\mu ^{\;\;\hat \alpha \hat \beta }&=& \tilde \omega _\mu ^{\;\;\hat \alpha \hat
    \beta     } \left[ {g } \right] - \frac{1}{3}\left( {N^{\hat \alpha \nu } N_{\;\;\mu} ^{\hat \beta }  - N_{\;\;\mu} ^{\hat \alpha } N^{\hat \beta \nu } } \right)\left( {\partial _\nu  \sigma } \right)\nonumber \\
    \omega _\mu ^{\;\;\hat a\hat b}  &=& W^{\hat a d} \left( {\partial _\mu
    W_{\;\;d}^{\hat b} } \right)  - \frac{1}{2}W^{\hat ad} W^{\hat bf} \left( {\partial _\mu  k_{df} } \right) \nonumber \\
    \omega _a^{\;\;\hat \alpha \hat b}  &=&  e^{ - \frac{\sigma }{2}} N^{\hat \alpha \beta
    }     \left[ {\frac{1}{6}W_{\;\;a}^{\hat b} \left( {\partial _\beta  \sigma } \right)-\frac{1}{2}W^{\hat b d} \left( {\partial _\beta  k_{ad} }
    \right)    }    \right]     \nonumber \\
    \omega _c^{\;\;\hat a\hat b}  &=& \hat \omega _c^{\;\;\hat a\hat b}\left[ {k_{ab} }
    \right].
 \end{eqnarray}

The spin connections carrying $\hat a$ and $\hat b$ indices break
down into $( {\hat m\hat {\bar n}} )$, $( {\hat {\bar m}\hat n}
)$, $\left( {\hat m\hat n} \right)$ and $( {\hat {\bar m}\hat
{\bar n}} )$ pieces. Based on the relations between the
deformations of the metric and the cohomology forms, such as
(\ref{modulideformations}), the non-vanishing ones can be written
in terms of variations of the moduli $z^i$. For example, one can
straightforwardly show that:
\begin{equation}\label{}
    \Gamma _{\mu m}^{\bar n}  = \frac{1}{{2\left\| \Omega  \right\|^2
    }}\Omega ^{\bar n\bar o\bar p} \chi_{\bar o\bar p m|\bar i} (
    {\partial _\mu  z{}^{\bar i}} ).
\end{equation}

To deal with the components $F_\mu$ of the field strength, we note
that, up to an exact form, one can always expand any three form in
terms of the ($3,0$) and ($2,1$) forms dual to the homology
decomposition (\ref{homologydecomposition}) as follows
\cite{Gutperle:2000ve,hep-th/9812049}:
\begin{eqnarray}
    F_\mu &=& ie^{{\K \mathord{\left/ {\vphantom {\K 2}} \right.
    \kern-\nulldelimiterspace} 2}} \bar B_\mu \Omega  - ie^{{\K\mathord{\left/{\vphantom {\K 2}} \right.
    \kern-\nulldelimiterspace} 2}} G^{i\bar j} \left( {\nabla _{\bar j} \bar B_\mu } \right)\left( {\nabla _i \Omega } \right) + c.c. \nonumber\\
    \bar B_\mu &=& \int {F_\mu   \wedge \bar \Omega },
\end{eqnarray}
where the quantities $B_\mu$ and $\bar B_\mu$ are the coefficients
of the expansion, found in the usual way by making use of
(\ref{Omegarelations}) and $c.c.$ represents the complex conjugate
of previous terms. The 3-form $F_\mu$ then becomes:
\begin{eqnarray}
    F_\mu &=& i\sqrt 2 \left[ {M_I ( {\partial _\mu  \zeta ^I }
    ) + L^I ( {\partial _\mu  \tilde \zeta _I } )}
    \right]\bar \Omega  \nonumber \\
    &-& i\sqrt 2 G^{i\bar j} \left[ {h_{iI} (
    {\partial _\mu  \zeta ^I } ) + f_i^I ( {\partial _\mu
    \tilde \zeta _I } )} \right]\left( {\nabla _{\bar j} \bar
    \Omega } \right) + c.c. \label{Fmu}
\end{eqnarray}

Putting everything together, we find that we can write the
resulting $D=5$ equations as follows: The gravitini variations:
\begin{eqnarray}
    \delta_\epsilon \psi  ^A  &=& \tilde{\nabla}  \epsilon^A + \left[ {\mathcal{G}  } \right]_{\;\;B}^A \epsilon ^B  \nonumber\\
    \left[ {\mathcal{G}  } \right] &=& \left[ {\begin{array}{*{20}c}
    {\frac{1}{4}\left( {v   - \bar v   - Y  } \right)} & { - \bar
    u     }  \\
    {u     } & { - \frac{1}{4}\left( {v   - \bar v   - Y  } \right)}
    \\
    \end{array}} \right] \nonumber \\ \label{gravitinotrans}
\end{eqnarray}
where the indices $A$ and $B$ run over $(1,2)$, $\tilde \nabla$ is given
by
\be
    \tilde{\nabla}=dx^\mu\left( \partial _\mu   + \frac{1}{4}\omega _\mu^{\,\,\,\,\hat \mu\hat \nu} \Gamma _{\hat \mu\hat
    \nu}\right)
\ee
as usual and
\begin{eqnarray}
    u   &=&  e^{\frac{\sigma }{2}} \left( {M_I  {d    \zeta^I } + L^I  {d  \tilde \zeta _I } }    \right) \nonumber \\
    \bar u   &=& e^{\frac{\sigma }{2}} \left( {\bar M_I  {d \zeta ^I }  + \bar L^I  {d  \tilde \zeta _I }    } \right) \nonumber \\
    v   &=& \frac{1}{2} {d  \sigma }    + \frac{i}{2}e^{-\sigma}  \star F \nonumber \\
    \bar v   &=& \frac{1}{2} {d  \sigma }    - \frac{i}{2}e^{-\sigma}  \star F.\label{eqns5}
\end{eqnarray}

The quantity $Y$ is proportional to the $U\left(1\right)$
connection $\mathcal{P}$ defined by (\ref{U1connection2});
explicitly:
\begin{equation}
    Y   = \frac{{\bar Z^I N_{IJ}  {d  Z^J }  -
    Z^I N_{IJ}  {d  \bar Z^J } }}{{\bar Z^I N_{IJ} Z^J
    }},
\end{equation}
where, as before, $N_{IJ}  = {\mathop{\rm Im}\nolimits} \left(
{F_{IJ} } \right)$ encoding the dependence of $F_I$ on $Z^I$. The
matrix $\mathcal{G}$ is the $Sp(1)$ connection of the quaternionic
manifold described by the action\footnote{Recall from Berger's
list that a quaternionic manifold has $Sp(h_{2,1}+1)\otimes Sp(1)$
holonomy.}. One can derive $\mathcal{G}$ based on this alone with
no reference to the higher dimensional theory, as was done in
\cite{Ferrara:1989ik} for the four dimensional case. The hyperini
equations are:
\begin{eqnarray}
    \delta_\epsilon \xi _1^I  = e_{\;\;\mu} ^{1I} \Gamma ^\mu  \epsilon _1  - \bar e_{\;\;\mu}^{2I}
    \Gamma ^\mu  \epsilon _2  \nonumber \\
    \delta_\epsilon \xi _2^I  = e_{\;\;\mu} ^{2I} \Gamma ^\mu  \epsilon _1  + \bar e_{\;\;\mu}^{1I}
    \Gamma ^\mu  \epsilon _2, \label{hyperinotrans}
\end{eqnarray}
written in terms of the quantities:
\begin{eqnarray}
    e ^{1I}&=&e_{\;\;\mu} ^{1I}dx^\mu  = \left( {\begin{array}{*{20}c}
   {u  }  \\
   {E ^{\hat i} }  \\
    \end{array}} \right) \nonumber \\\nonumber \\
    e^{2I}&=&e_{\;\;\mu} ^{2I}dx^\mu  = \left(
    {\begin{array}{*{20}c}
   {v  }  \\
   {e ^{\hat i} }  \\
    \end{array}} \right)
\end{eqnarray}
\begin{eqnarray}
    E ^{\hat i}  &=&  e^{\frac{\sigma }{2}} e^{\hat ij} \left( {h_{jI}    {d  \zeta ^I }  + f_j^I  {d  \tilde \zeta _I }    } \right) \nonumber \\
    \bar E ^{\hat i}  &=&  e^{\frac{\sigma }{2}} e^{\hat i\bar j} \left( {h_{\bar    jI}      {d \zeta ^I }  + f_{\bar j}^I  {d  \tilde \zeta _I } } \right),
\end{eqnarray}
and the beins of the special K\"{a}hler metric:
\begin{eqnarray}
    e ^{\hat i}  &=& e_{\;\;j}^{\hat i}  {d  z^j } \quad,\quad \quad
    \quad \bar e^{\hat i}  = e_{\;\;{\bar j}}^{\hat i}  {d  z^{\bar j} } \nonumber \\
    G_{i\bar j}  &=& e_{\;\;i}^{\hat k} e_{\;\;{\bar j}}^{\hat l} \delta _{\hat k\hat l}.
\end{eqnarray}

These quantities may also be used to make the connection between the special K\"{a}hler language we are using here and the quaternionic language used more abundantly in the literature. Quaternionic vielbeins may be defined as follows:
\begin{equation}
V^{\Gamma A}  = \left( {\begin{array}{*{20}c}
   {e ^{1I} }  \\
   {\bar e ^{2I} }  \\
   { - e ^{2I} }  \\
   {\bar e ^{1I} }  \\
\end{array}} \right),\quad \quad \Gamma  = 1, \ldots ,2\left(h_{2,1}+1\right) ,\quad A = 1,2
\end{equation}
such that:
\bea
    \int {h_{uv} dq^u  \wedge \star dq^v }
    &=& 2\int { \left( {u \wedge \star  \bar u   + v \wedge \star \bar v   + \delta _ {\hat i \hat j} e^{\hat i}\wedge \star \bar e^{\hat j}  + \delta _ {\hat i \hat j} E^{\hat i}\wedge \star \bar E^{\hat j} }
    \right)},
\eea
where $h_{uv}$ is the quaternionic metric with coordinates $q^u$; the hypermultiplet scalars. This is tantamount to demonstrating the c-map, which relates the quaternionic form of the hypermultiplets in $D=5$ to the \textbf{SKG} form of the vector multiplets in $D=4$. The proof that this is, in fact, a quaternionic structure as defined in \S\ref{Quaternionic Manifolds} is somewhat tedious. The interested reader may consult \cite{Ferrara:1989ik}.

\subsection{The theory in manifestly symplectic form}

For the sake of completeness, we also give a recently proposed form of the $\N=2$ theory \cite{Emam:2009xj}, clearly highlighting its symplectic structure. Since the action is invariant under rotations in \textbf{\textit{Sp}},
then it is clear that $R$, $d\sigma$, $dz$ and $F$
are themselves symplectic invariants. The axion fields
$\left(\zeta, \tilde\zeta\right)$, however, can be thought of as
components of an \textbf{\textit{Sp}} `axions vector'. If we
define:
\be\label{XiasSp}
   \left| \Xi  \right\rangle  = \left( {\begin{array}{*{20}c}
   {\,\,\,\,\,\zeta ^I }  \\
   -{\tilde \zeta _I }  \\
    \end{array}} \right), \quad\quad\quad\quad\left| {d\Xi } \right\rangle  = \left( {\begin{array}{*{20}c}
   {\,\,\,\,\,d\zeta ^I }  \\
   -{d\tilde \zeta _I }  \\
    \end{array}} \right)
\ee
then
\be
    \left\langle {{\Xi }}
 \mathrel{\left | {\vphantom {{\Xi } d\Xi }}
 \right. \kern-\nulldelimiterspace}
 {d\Xi } \right\rangle   = \zeta^I d\tilde \zeta_I  - \tilde \zeta_I
 d\zeta^I,
\ee
as well as:
\bea
    \lefteqn{
    \left\langle {\partial _\mu  \Xi } \right|\Lambda \left| {\partial ^\mu  \Xi }
    \right\rangle
    }\nonumber\\
    &=&    -\left( {\gamma_{IJ}  + \gamma ^{ KL} \theta _{IK}\theta _{JL} } \right) \left( {\partial _\mu  \zeta ^I } \right) \left( {\partial ^\mu  \zeta ^J } \right)  - \gamma ^{IJ}  \left( {\partial _\mu  \tilde \zeta _I } \right) \left( {\partial ^\mu  \tilde \zeta _J } \right)  - 2\gamma ^{ IK} \theta_{JK}  \left( {\partial _\mu  \zeta ^J } \right) \left( {\partial ^\mu  \tilde \zeta _I }
    \right),\nonumber\\
\eea
such that (\ref{X}) becomes
\bea
    X&=& {\left( {\gamma_{IJ}  + \gamma ^{ KL} \theta _{IK}\theta _{JL} } \right) {d\zeta^I } \wedge\star {d\zeta^J }  + \gamma ^{IJ}  {d\tilde \zeta_I } \wedge\star {d\tilde \zeta_J }  + 2\gamma ^{ IK} \theta_{JK}  {d\zeta^J } \wedge\star {d\tilde \zeta_I } }
   \nonumber\\&=&  - \left\langle {\partial _\mu  \Xi } \right|\Lambda \left| {\partial ^\mu  \Xi } \right\rangle \star   1.
\eea

As a consequence of this language, the field expansion
(\ref{FExpanded}) could be rewritten
\bea
    \A &=& A + \sqrt 2  \left\langle {\Theta }
 \mathrel{\left | {\vphantom {\Theta  \Xi }}
 \right. \kern-\nulldelimiterspace}
 {\Xi } \right\rangle,\nonumber\\
    \F &=& d\A = F + \sqrt 2 \mathop {\left\langle {\Theta }
 \mathrel{\left | {\vphantom {\Theta  {d\Xi }}}
 \right. \kern-\nulldelimiterspace}
 {{d\Xi }} \right\rangle }\limits_{ \wedge \,\,\,\,}. \label{FExpandedSp}
\eea

The bosonic action in manifest symplectic covariance is hence:
\bea
    S_5  &=& \int\limits_5 {\left[ {R\star 1 - \frac{1}{2}d\sigma \wedge\star d\sigma  - G_{i\bar j} dz^i \wedge\star dz^{\bar j} } \right.}  \nonumber\\
    & &\left. {- F \wedge \left\langle {{\Xi }}
    \mathrel{\left | {\vphantom {{\Xi } d\Xi }} \right. \kern-\nulldelimiterspace} {d\Xi } \right\rangle  - \frac{1}{2}e^{ - 2\sigma } F \wedge \star F
    + e^\sigma   \left\langle {\partial _\mu  \Xi } \right|\Lambda \left| {\partial ^\mu  \Xi } \right\rangle\star 1}
    \right].
\eea

The equations of motion are now
\bea
    \left( {\Delta \sigma } \right)\star 1 + e^\sigma   \left\langle {\partial _\mu  \Xi } \right|\Lambda \left| {\partial ^\mu  \Xi } \right\rangle \star 1 + e^{ - 2\sigma } F \wedge \star F &=& 0\label{dilatoneomSp}\\
    \left( {\Delta z^i } \right)\star 1 + \Gamma _{jk}^i dz^j  \wedge \star dz^k  + \frac{1}{2}e^\sigma  G^{i\bar j}  {\partial _{\bar j} \left\langle {\partial
    _\mu \Xi}    \right|\Lambda \left| {\partial ^\mu  \Xi }   \right\rangle\star  1} &=& 0 \nonumber\\
    \left( {\Delta z^{\bar i} } \right)\star 1 + \Gamma _{\bar j\bar k}^{\bar i} dz^{\bar j}  \wedge \star dz^{\bar k}  + \frac{1}{2}e^\sigma  G^{\bar ij}  {\partial _j \left\langle {\partial _\mu
    \Xi}   \right|\Lambda \left| {\partial ^\mu  \Xi } \right\rangle\star 1}  &=& 0\label{zeomSp} \\
    d^{\dag} \left[ {e^{ - 2\sigma } F + \star\left\langle {{\Xi }} \mathrel{\left | {\vphantom {{\Xi } d\Xi }}
     \right. \kern-\nulldelimiterspace} {d\Xi } \right\rangle} \right] &=& 0\label{FeomgeneralSp}\\
    d^\dag\left[ {e^\sigma  \left| {\Lambda d\Xi } \right\rangle  +  \star F \left| {\Xi } \right\rangle } \right] &=&0.\label{xieomSp}
\eea

Note that, as is usual for Chern-Simons actions, the explicit
appearance of the gauge potential $\left| \Xi  \right\rangle $ in
(\ref{FeomgeneralSp}) and (\ref{xieomSp}) does not have an effect
on the physics since:
\bea
    d^{\dag} \star\left\langle {\Xi }
    \mathrel{\left | {\vphantom {\Xi  {d\Xi }}}
    \right. \kern-\nulldelimiterspace}
    {{d\Xi }} \right\rangle  & &\longrightarrow\quad d\left\langle {\Xi }
    \mathrel{\left | {\vphantom {\Xi  {d\Xi }}}
    \right. \kern-\nulldelimiterspace}
    {{d\Xi }} \right\rangle  = \mathop {\left\langle {{d\Xi }}
    \mathrel{\left | {\vphantom {{d\Xi } {d\Xi }}}
    \right. \kern-\nulldelimiterspace}
    {{d\Xi }} \right\rangle }\limits_ \wedge\nonumber\\
    d^{\dag} \star F\left| \Xi  \right\rangle  & &\longrightarrow\quad d\left[ {F\left| \Xi  \right\rangle } \right] = F\wedge \left| {d\Xi } \right\rangle,
\eea
where the Bianchi identities on $A$ and ${\left| \Xi \right\rangle
}$ were used. The Noether currents and
charges become
\bea
    \mathcal{J}_2  &=& e^{ - 2\sigma } F + \star\left\langle {{\Xi }} \mathrel{\left | {\vphantom {{\Xi } d\Xi }}
     \right. \kern-\nulldelimiterspace} {d\Xi } \right\rangle\nonumber\\
     \left| {\mathcal{J}_5 } \right\rangle &=&e^\sigma  \left| {\Lambda d\Xi } \right\rangle  +  \star F \left| {\Xi } \right\rangle\nonumber\\
    \mathcal{Q}_2  &=& \int {\mathcal{J}_2 },\quad \quad \quad
    \left| {\mathcal{Q}_5 } \right\rangle  = \int {\left| {\mathcal{J}_5 } \right\rangle }.
    \label{CurrentsChargesSp}
\eea

The equations of the universal axion (\ref{UniversalAxion}),
(\ref{a field equation}) and (\ref{a action}) are now
\be
    da = e^{ - 2\sigma } \star F - \left\langle {\Xi } \mathrel{\left | {\vphantom {\Xi  {d\Xi }}} \right. \kern-\nulldelimiterspace} {{d\Xi }}
    \right\rangle,
\ee
\be
    d^{\dag} \left[ {e^{2\sigma } da + e^{2\sigma } \left\langle {\Xi } \mathrel{\left | {\vphantom {\Xi  {d\Xi }}} \right. \kern-\nulldelimiterspace} {{d\Xi }}
    \right\rangle} \right] =    0\quad\quad {\rm and}
\ee
\be
    S_a  = \frac{1}{2}\int {e^{2\sigma } \left[ {da + \left\langle {\Xi } \mathrel{\left | {\vphantom {\Xi  {d\Xi }}} \right. \kern-\nulldelimiterspace} {{d\Xi }}
    \right\rangle} \right] \wedge \star\left[ {da + \left\langle {\Xi } \mathrel{\left | {\vphantom {\Xi  {d\Xi }}} \right. \kern-\nulldelimiterspace} {{d\Xi }}
    \right\rangle}    \right]}.
\ee

The gravitini equations can
be explicitly written as follows:
\bea
 \delta _\epsilon  \psi ^1  &=& \tilde{\nabla} \epsilon _1  + \frac{1}{4}\left( {ie^{ - \sigma } \star F - Y} \right)\epsilon _1  - e^{\frac{\sigma }{2}} \left\langle {{\bar V}}
 \mathrel{\left | {\vphantom {{\bar V} {d\Xi }}} \right. \kern-\nulldelimiterspace} {{d\Xi }} \right\rangle\epsilon _2  \\
 \delta _\epsilon  \psi ^2  &=& \tilde{\nabla} \epsilon _2  - \frac{1}{4}\left( {ie^{ - \sigma } \star F - Y} \right)\epsilon _2  + e^{\frac{\sigma }{2}} \left\langle {V}
 \mathrel{\left | {\vphantom {V {d\Xi }}} \right. \kern-\nulldelimiterspace} {{d\Xi }} \right\rangle \epsilon _1,\label{SUSYSpGravitini}
\eea
while the hyperini variations are
\bea
     \delta _\epsilon  \xi _1^0  &=& e^{\frac{\sigma }{2}} \left\langle {V}
    \mathrel{\left | {\vphantom {V {\partial _\mu  \Xi }}} \right. \kern-\nulldelimiterspace} {{\partial _\mu  \Xi }} \right\rangle  \Gamma ^\mu  \epsilon _1  - \left[ {\frac{1}{2}\left( {\partial _\mu  \sigma } \right) - \frac{i}{2}e^{ - \sigma } \left( {\star F} \right)_\mu  } \right]\Gamma ^\mu  \epsilon _2  \nonumber\\
     \delta _\epsilon  \xi _2^0  &=& e^{\frac{\sigma }{2}} \left\langle {{\bar V}}
    \mathrel{\left | {\vphantom {{\bar V} {\partial _\mu  \Xi }}} \right. \kern-\nulldelimiterspace} {{\partial _\mu  \Xi }} \right\rangle \Gamma ^\mu  \epsilon _2  + \left[ {\frac{1}{2}\left( {\partial _\mu  \sigma } \right) + \frac{i}{2}e^{ - \sigma } \left( {\star F} \right)_\mu  } \right]\Gamma ^\mu  \epsilon
     _1\label{SUSYSpHyperiniFirst}\\
     \delta _\epsilon  \xi _1^{\hat i}  &=& e^{\frac{\sigma }{2}} e^{\hat ij} \left\langle {{U_j }}
    \mathrel{\left | {\vphantom {{U_j } {\partial _\mu  \Xi }}} \right. \kern-\nulldelimiterspace} {{\partial _\mu  \Xi }} \right\rangle \Gamma ^\mu  \epsilon _1  - e_{\,\,\,\bar j}^{\hat i} \left( {\partial _\mu  z^{\bar j} } \right)\Gamma ^\mu  \epsilon _2  \nonumber\\
     \delta _\epsilon  \xi _2^{\hat i}  &=& e^{\frac{\sigma }{2}} e^{\hat i\bar j} \left\langle {{U_{\bar j} }}
    \mathrel{\left | {\vphantom {{U_{\bar j} } {\partial _\mu  \Xi }}} \right. \kern-\nulldelimiterspace} {{\partial _\mu  \Xi }} \right\rangle \Gamma ^\mu  \epsilon _2  + e_{\,\,\,j}^{\hat i} \left( {\partial _\mu  z^j } \right)\Gamma ^\mu  \epsilon
     _1.\label{SUSYSpHyperini}
\eea

Finally a useful set of identities was derived in \cite{Emam:2009xj} which we reproduce here for easy reference:
\bea
    dG_{i\bar j}  &=& G_{k\bar j} \Gamma _{ri}^k dz^r  + G_{i\bar k} \Gamma _{\bar r\bar j}^{\bar k} dz^{\bar r}  \nonumber\\
    dG^{i\bar j}  &=&  - G^{p\bar j} \Gamma _{rp}^i dz^r  - G^{i\bar p} \Gamma _{\bar r\bar p}^{\bar j} dz^{\bar r}  \nonumber\\
    \left| {dV} \right\rangle  &=& dz^i \left| {U_i } \right\rangle  - i\mathcal{P}\left| V \right\rangle \nonumber \\
    \left| {d\bar V} \right\rangle  &=& dz^{\bar i} \left| {U_{\bar i} } \right\rangle  + i\mathcal{P}\left| {\bar V} \right\rangle \nonumber \\
    \left| {dU_i } \right\rangle  &=& G_{i\bar j} dz^{\bar j} \left| V \right\rangle  + \Gamma _{ik}^r dz^k \left| {U_r } \right\rangle+G^{j\bar l} C_{ijk} dz^k \left| {U_{\bar l} } \right\rangle - i\mathcal{P}\left| {U_i } \right\rangle \nonumber \\
    \left| {dU_{\bar i} } \right\rangle  &=& G_{j\bar i} dz^j \left| {\bar V} \right\rangle + \Gamma _{\bar i\bar k}^{\bar r} dz^{\bar k} \left| {U_{\bar r} } \right\rangle + G^{l\bar j} C_{\bar i\bar j\bar k} dz^{\bar k} \left| {U_l } \right\rangle + i\mathcal{P}\left| {U_{\bar i} } \right\rangle \nonumber \\
    d{\bf \Lambda } &=& \left( {\partial _i {\bf \Lambda }} \right)dz^i  + \left( {\partial _{\bar i} {\bf \Lambda }} \right)dz^{\bar i},\label{SpacetimeVariations}
\eea
where $\mathcal{P}$ is the $U\left(1\right)$ connection defined by
(\ref{U1connection2}):
\be
    \mathcal{P} =  - \frac{i}{2}\left[ {\left( {\partial _i \K} \right)dz^i  - \left( {\partial _{\bar i} \K} \right)dz^{\bar i} }
    \right],
\ee
and $\left( {\partial _i {\bf \Lambda }} , {\partial _{\bar i}
{\bf \Lambda }}\right)$ are given by (\ref{CovariantDerofLambda}).

\pagebreak

\renewcommand{\theequation}{A-\arabic{equation}}
  \setcounter{equation}{0}  
  \section*{Appendix: Differential forms on manifolds}  

In this appendix we review the language of differential forms used
in various locations in the text. Clearly, reading this review
requires more knowledge of differential forms, Hodge theory and
topology than is reviewed here. The purpose of this appendix is
then to simply set the notation and collect in one place all the
equations necessary to reproduce the various details in the review.

Consider a $D$-dimensional Riemannian/Lorentzian manifold $\M$. A
differential form $\omega$ or $\omega_p$ of order $p$ on $\M$,
also known as a $p$-form, is a totally antisymmetric tensor of
type $\left(0,p\right)$. It may be defined in terms of the
differentials $dx^\mu$, themselves 1-forms, acting as basis in
this case, in the standard way:
\be
    \omega  = \frac{1}{{p!}}\omega _{\mu _1  \cdots \mu _p } dx^{\mu _1 }  \wedge  \cdots  \wedge dx^{\mu_p},
\ee
where the so-called \textbf{wedge product} $\wedge$ is defined
such that the following properties are satisfied
\begin{enumerate}
    \item $dx^{\mu _1 }  \wedge  \cdots  \wedge dx^{\mu _p }  = 0$ if some index $\mu _i$ appears at least twice.
    \item $dx^{\mu _1 }  \wedge  \cdots  \wedge dx^{\mu _p }  =  - dx^{\mu _1 }  \wedge  \cdots  \wedge dx^{\mu _p }$ on the exchange of two adjacent
    indices.
    \item $dx^{\mu _1 }  \wedge  \cdots  \wedge dx^{\mu _p }$ is linear in each $dx^{\mu _i}$.
\end{enumerate}

Clearly the components $\omega _{\mu _1  \cdots \mu _p }$ are
themselves antisymmetric such that $\omega$ is nonvanishing. It
can be shown that the wedge product of forms $\omega_p$, $\eta_r$
and $\xi_q$ satisfies
\begin{enumerate}
    \item $\omega _p  \wedge \omega _p  = 0$ if $p$ is odd.
    \item $\omega _p  \wedge \eta _r  = \left( { - 1} \right)^{pr} \eta _r  \wedge \omega
    _p$.
    \item $\left( {\xi  \wedge \omega } \right) \wedge \eta  = \xi  \wedge \left( {\omega  \wedge \eta }
    \right)$.
\end{enumerate}

The so-called \textbf{exterior derivative} $d = dx^\nu  \partial
_\nu  $ is defined as an operator that maps $p$-forms into
$\left(p+1\right)$-forms as follows
\be
    \lambda _{p + 1}  = d\omega _p  = dx^\nu {\partial _\nu  \omega _p }    = \frac{1}{{p!}}\left( {\partial _\nu  \omega _{\mu _1  \cdots \mu _p } } \right)dx^\nu   \wedge dx^{\mu _1 }  \wedge  \cdots  \wedge dx^{\mu _p
    },
\ee
satisfying the product rule
\be
    d\left( {\omega _p  \wedge \eta _r } \right) = \left( {d\omega _p } \right) \wedge \eta _r  + \left( { - 1} \right)^p \omega _p  \wedge \left( {d\eta _r }
    \right),
\ee
as well as the very important
\be\label{nilpotent}
    d^2  = 0.
\ee

An operator satisfying (\ref{nilpotent}) is called
\textbf{nilpotent}. Differential forms that can be written as
exterior derivatives of other forms, such as $\lambda=d\omega$,
are called \textbf{exact}, while forms whose exterior derivative
vanishes, \emph{e.g.} $d\lambda=0$, are called \textbf{closed}.
Because of (\ref{nilpotent}), exact forms are always closed, while
the converse is not necessarily true.

One of the most beautiful theorems in mathematics involves the
integration of $p$-forms and is known as \textbf{Stokes' theorem},
which is a generalization of the familiar theorem by the same name
in $\mathbb{R}^3$, as well as the divergence theorem and the
fundamental theorem of calculus. It states
\be\label{stokes}
    \int\limits_\M {d\omega }  = \int\limits_{\partial \M} \omega,
\ee
where $\partial \M$ denotes the $\left(D-1\right)$-dimensional
boundary of $\M$, unless of course $\M$ is closed, in which case
$\omega$ is closed as well and the right hand side vanishes.
Equation (\ref{stokes}) is sometimes referred to as the
fundamental theorem of calculus on manifolds.

Note that all our definitions so far are metric independent. We
now choose a metric $g_{\mu\nu}$ on $\M$ with either Riemannian or
Lorentzian signatures. We also define the Levi-Civita totally
antisymmetric symbol in the following way:
\be
    \bar \varepsilon _{\mu _1  \cdots \mu _p }  = \left\{ {\begin{array}{*{20}c}
   { + 1} & {{\text{for even permutations of the indices.}}}  \\
   { - 1} & {{\text{for odd permutations of the indices.}}}  \\
   {\;\;\rm 0} & {{\text{if some index } }\mu _i {\text{ appears at least twice.}}}  \\
    \end{array}} \right.
\ee
where $\bar\varepsilon_{0  \cdots D-1 }\text{ or
}\bar\varepsilon_{1 \cdots D}=+1$. Defined this way, $\bar
\varepsilon _{\mu _1 \cdots \mu _p }$ does not transform as a
tensor, hence the name `symbol'. One way of defining a Levi-Civita
\emph{tensor} is described below. The volume form over $D$
dimensions is defined by
\be\label{VolumeForm}
    \varepsilon _D  =\sqrt {\left| g \right|} dx^1  \wedge  \cdots  \wedge dx^D  =
    \frac{1}{{D!}}\varepsilon _{\mu _1  \cdots \mu _D } dx^{\mu _1 }
    \wedge  \cdots  \wedge dx^{\mu _D },
\ee
where the unbarred $\varepsilon _{\mu _1  \cdots \mu _D }$ (the
components of $\varepsilon _D$, which does transform as a tensor)
are defined by
\bea
    \varepsilon _{\mu _1  \cdots \mu _D }  &=& \sqrt {\left| g \right|} \bar \varepsilon _{\mu _1  \cdots \mu _D }  \nonumber\\
    \varepsilon ^{\mu _1  \cdots \mu _D }  &=& \frac{1}{{\sqrt {\left| g \right|} }}\bar \varepsilon ^{\mu _1  \cdots \mu
    _D    }.
\eea

The indices of $\varepsilon _{\mu _1 \cdots \mu _D }$ are raised
and lowered by $g_{\mu\nu}$ while those of $\bar \varepsilon _{\mu
_1 \cdots \mu _D }$ are raised and lowered by the flat metric
(either Minkowski or Euclidean depending on the signature of
$g_{\mu\nu}$). Clearly:
\be
    dx^{\mu _1 } \wedge  \cdots  \wedge dx^{\mu _D }= \bar \varepsilon^{\mu _1  \cdots \mu _D
    }dx^1  \wedge  \cdots  \wedge dx^D.
\ee

Note that in most of the literature the
nomenclature $\sqrt {\left| g \right|} d^D x$ is used as a
substitute for $\sqrt {\left| g \right|} dx^1 \wedge \cdots \wedge
dx^D$ which is technically the correct volume element.

Based on all this, we define the \textbf{Hodge-duality} operator
$\star$, mapping $p$-forms into $\left(D-p\right)$-forms, as
follows
\bea
 \star\left( {dx^{\mu _1 }  \wedge  \cdots  \wedge dx^{\mu _p } } \right) &=& \frac{1}{{\left( {D - p} \right)!}}\varepsilon _{\,\,\,\,\,\,\,\,\,\,\,\,\,\,\,\,\,\,\,\mu _{p + 1}  \cdots \mu _D }^{\mu _1  \cdots \mu _p } dx^{\mu _{p + 1} }  \wedge  \cdots  \wedge dx^{\mu _D }  \nonumber\\
 \kappa _{D - p}  = \star\omega _p  &=& \frac{1}{{p!\left( {D - p} \right)!}}\varepsilon _{\mu _1  \cdots \mu _p \mu _{p + 1}  \cdots \mu _D } \omega ^{\mu _1  \cdots \mu _p } dx^{\mu _{p + 1} }  \wedge  \cdots  \wedge dx^{\mu _D }  \nonumber\\
  &=& \frac{1}{{\left( {D - p} \right)!}}\kappa _{\mu _{p + 1}  \cdots \mu _D } dx^{\mu _{p + 1} }  \wedge  \cdots  \wedge dx^{\mu _D
  }.
\eea

Note that, in this language, the volume form is the Hodge dual of
the identity, \emph{i.e.}
\be
    \star 1 = \sqrt {\left| g \right|} dx^1  \wedge  \cdots  \wedge
    dx^D.
\ee

Furthermore, it can be straightforwardly shown that
\bea
    \star\star\omega_p  &=& \left( { - 1} \right)^{p\left( {D - p} \right) + \varrho}
    \omega_p\nonumber\\
    \star^{ - 1}  &=& \left( { - 1} \right)^{p\left( {D - p} \right) + \varrho}\star,
\eea
where $\star^{-1}$ is the inverse Hodge dual and $\varrho$ is the
number of eigenvalues of the metric with a minus sign, \emph{i.e.}
if $\M$ is Riemannian then $\varrho=0$, while if it is Lorentzian
then $\varrho=1$.

We can now define an inner product of forms. This is
\be
    \left( {\omega _p ,\eta _p } \right) = \int\limits_\M {\omega _p  \wedge \star\eta _p
    },
\ee
where
\be
    \omega _p  \wedge \star\eta _p  = \frac{1}{{p!}}\omega _{\mu _1  \cdots \mu _p } \eta ^{\mu _1  \cdots \mu _p } \sqrt {\left| g \right|} dx^1  \wedge  \cdots  \wedge
    dx^D.
\ee

The inner product is clearly symmetrical:
\be
   \begin{array}{*{20}c}
   \because & {\omega _p  \wedge \star\eta _p  = \eta _p  \wedge \star\omega _p }  \\
   \therefore & {\left( {\omega _p ,\eta _p } \right) \,\,= \left( {\eta _p ,\omega _p } \right).}  \\
    \end{array}
\ee

We can also use the Hodge dual to define the so-called
\textbf{adjoint exterior derivative operator}:
\be
    d^\dag  \omega_p= \left( { - 1} \right)^{D\left( {p + 1} \right) + \varrho}
    \star d\star \omega_p,
\ee
which maps a $p$-form \emph{down} to a $\left(p-1\right)$-form:
\be
    \star d\star\omega _p  = \star d\kappa _{D - p}  = \star\tau _{D - p + 1}  = \phi _{D - \left( {D - p + 1} \right)}  = \phi _{p -    1}.
\ee

Also note that since $d$ is nilpotent, then so is $d^\dag$:
\be
    {d^{\dag}} ^2  \propto \star d\star\star d\star \propto \star d^2 \star = 0.
\ee

For calculational convenience, we explicitly give the action of
the adjoint exterior derivative on a $p$-form $\omega$ in $D$
dimensions:
\be\label{d Dagger}
    d^{\dag} \omega  = \frac{{\left( { - 1} \right)^{\left(D+1\right)\left(p+1\right) }
    }}{{p!\left( {p - 1} \right)!}}\left( {\nabla ^{\mu _1 } \omega
    _{\mu _1\mu_2 \cdots \mu _p } } \right)dx^{\mu _2 }  \wedge  \cdots
    \wedge dx^{\mu _p },
\ee
where $\nabla_\mu$ is the usual Levi-Civita connection with
respect to the metric on $\M$. Certain useful theorems involving
$d^\dag$ can be proven. For example, one can show that
\be
    \left( {d\omega ,\eta } \right) = \left( {\omega ,d^{\dag} \eta }
    \right).
\ee

In analogy with the exterior derivative, one says that a form
$\lambda$ that can be written as $\lambda=d^\dag \omega$ is
\textbf{co-exact}, while one that satisfies $d^\dag\lambda=0$ is
\textbf{co-closed}. Clearly, co-exact forms are always co-closed
while the converse is not necessarily true.

Finally, we define the \textbf{Laplacian operator} on $p$-forms
by\footnote{Also sometimes known as the \textbf{Laplace-de Rahm
operator}, to differentiate between it and the ordinary Laplacian
$\nabla^2$ acting on scalar functions, which is a special case as
we will see.}
\be
    \Delta  = \left( {d + d^{\dag} } \right)^2  = dd^{\dag}  + d^{\dag}
    d.
\ee

A $p$-form that satisfies
\be\label{HarmonicCondition}
    \Delta\omega=0
\ee
is called \textbf{harmonic}, and (\ref{HarmonicCondition}) is
known as the \textbf{harmonic condition}. A form is harmonic if
and only if it is both closed \emph{and} co-closed. Harmonic forms
clearly play a fundamental role in physics. To demonstrate,
consider the following: Using (\ref{d Dagger}), the Laplacian of a
$0$-form scalar field $f$, \emph{i.e.} an ordinary function, in
$D=5$ spacetime, leads to the familiar:
\be
    \Delta f = d^{\dag} d f =  \nabla ^\mu  \nabla _\mu  f = \nabla ^2
    f,
\ee
where we have used $d^\dag f=0$. Also consider the Laplacian for a
general Abelian gauge potential $A$ in $D$ dimensions. We define $F=dA$ as
usual and write:
\be
    \Delta A = dd^{\dag} A + d^{\dag} d A = dd^{\dag} A + d^{\dag} F.
\ee

Because of the gauge freedom of $A$, we normally choose $d^{\dag}
A=0$, which is the generalized \textbf{Lorenz gauge
condition}\footnote{Note that this refers to the Danish physicist
L. Lorenz and not the Dutch H. Lorentz, of Lorentz transformations
fame. Confusing the two names is a recurring error in the literature.} leading to the
more familiar $\nabla _{\mu _1 } A^{\mu _1 \mu _2  \cdots }  = 0$.
Hence
\be
    \Delta A = d^{\dag} d A = d^{\dag} F,
\ee
which, in physical theory, may or may not vanish depending on the
presence or absence of sources. For example, in ordinary Maxwell
theory in $D=4$ flat spacetime, the expression $d^{\dag} F=J$
leads to the ordinary Gauss and Amp\`{e}re laws, provided that $J$
is the current 1-form. The Bianchi identity $dF=0$, resulting from
the fact that the $U\left(1\right)$ form $F$ is exact, leads to
the Faraday and no-monopoles laws.

It is clearly straightforward to extend the formalism of
differential forms to complex manifolds. We will not do so here
but rather refer the interested reader to more detailed
discussions of this vast topic, such as \cite{Nakahara:2003nw}.

\pagebreak


\begin{thebibliography}{999}

\bibitem{Susskind:2005bd}
  L.~Susskind,
{\it  New York, USA: Little, Brown (2005) 403 p}

\bibitem{Freivogel:2005vv}
  B.~Freivogel, M.~Kleban, M.~Rodriguez Martinez and L.~Susskind,
  JHEP {\bf 0603}, 039 (2006)
  [arXiv:hep-th/0505232].

\bibitem{Freivogel:2004rd}
  B.~Freivogel and L.~Susskind,
  Phys.\ Rev.\  D {\bf 70}, 126007 (2004)
  [arXiv:hep-th/0408133].

\bibitem{Susskind:2004uv}
  L.~Susskind,
  arXiv:hep-th/0405189.

\bibitem{Susskind:2003kw}
  L.~Susskind,
  arXiv:hep-th/0302219.

\bibitem{Smolin:2006pe}
  L.~Smolin,
{\it  Boston, USA: Houghton Mifflin (2006) 392 p}

\bibitem{Woit:2006js}
  P.~Woit,
{\it  London, UK: Cape (2006) 290 p}

\bibitem{Emam:2008js}
  M.~H.~Emam,
  Am.\ J.\ Phys.\  {\bf 76}, 605 (2008)
  [arXiv:0805.0543 [physics.pop-ph]].

\bibitem{Maldacena:2001uc}
  J.~M.~Maldacena,
  ``Large N field theories, string theory and gravity,''
    {\it Prepared for ICTP Spring School on Superstrings and
    Related Matters, Trieste, Italy, 2-10 Apr 2001}.

\bibitem{hep-th/9508072}
  S.~Ferrara, R.~Kallosh and A.~Strominger,
  ``N=2 extremal black holes,''
  Phys.\ Rev.\  D {\bf 52}, 5412 (1995)
  [arXiv:hep-th/9508072].

\bibitem{hep-th/9602111}
  A.~Strominger,
  ``Macroscopic Entropy of $N=2$ Extremal Black Holes,''
  Phys.\ Lett.\  B {\bf 383}, 39 (1996)
  [arXiv:hep-th/9602111].

\bibitem{hep-th/9602136}
  S.~Ferrara and R.~Kallosh,
  ``Supersymmetry and Attractors,''
  Phys.\ Rev.\  D {\bf 54}, 1514 (1996)
  [arXiv:hep-th/9602136].

\bibitem{hep-th/0506177}
  A.~Sen,
  ``Black hole entropy function and the attractor mechanism in higher derivative gravity,''
  JHEP {\bf 0509}, 038 (2005)
  [arXiv:hep-th/0506177].

\bibitem{hep-th/0507096}
  K.~Goldstein, N.~Iizuka, R.~P.~Jena and S.~P.~Trivedi,
  ``Non-supersymmetric attractors,''
  Phys.\ Rev.\  D {\bf 72}, 124021 (2005)
  [arXiv:hep-th/0507096].

\bibitem{hep-th/0508042}
  A.~Sen,
  ``Entropy function for heterotic black holes,''
  JHEP {\bf 0603}, 008 (2006)
  [arXiv:hep-th/0508042].

\bibitem{Cho:2000hg}
  H.~Cho, M.~Emam, D.~Kastor and J.~H.~Traschen,
  ``Calibrations and Fayyazuddin-Smith spacetimes,''
  Phys.\ Rev.\  D {\bf 63}, 064003 (2001)
  [arXiv:hep-th/0009062].

\bibitem{Kastor:2003jy}
  D.~Kastor,
  ``From wrapped M-branes to Calabi-Yau black holes and strings,''
  JHEP {\bf 0307}, 040 (2003)
  [arXiv:hep-th/0305261].

\bibitem{Martelli:2003ki}
  D.~Martelli and J.~Sparks,
  ``G-structures, fluxes and calibrations in M-theory,''
  Phys.\ Rev.\  D {\bf 68}, 085014 (2003)
  [arXiv:hep-th/0306225].

\bibitem{Fayyazuddin:2005as}
  A.~Fayyazuddin and T.~Z.~Husain,
  ``The geometry of M-branes wrapping special Lagrangian cycles,''
  Class.\ Quant.\ Grav.\  {\bf 23}, 7245 (2006)
  [arXiv:hep-th/0505182].

\bibitem{Emam:2005bh}
  M.~H.~Emam,
  ``Five dimensional 2-branes from special Lagrangian wrapped M5-branes,''
  Phys.\ Rev.\  D {\bf 71}, 125020 (2005)
  [arXiv:hep-th/0502112].

\bibitem{Emam:2006sr}
  M.~H.~Emam,
  ``Wrapped M5-branes leading to five dimensional 2-branes,''
  Phys.\ Rev.\  D {\bf 74}, 125004 (2006)
  [arXiv:hep-th/0610161].

\bibitem{Emam:2007qa}
  M.~H.~Emam,
  ``Five dimensional 2-branes and the universal hypermultiplet,'' Nuclear Physics B (2009), doi:10.1016/j.nuclphysb.2009.02.012 [arXiv:hep-th/0701060].

\bibitem{Ferrara:1989ik}
  S.~Ferrara and S.~Sabharwal,
  ``Quaternionic Manifolds for Type II Superstring Vacua of Calabi-Yau Spaces,''
  Nucl.\ Phys.\  B {\bf 332}, 317 (1990).

\bibitem{Gutperle:2000ve}
  M.~Gutperle and M.~Spalinski,
  ``Supergravity instantons for N = 2 hypermultiplets,''
  Nucl.\ Phys.\  B {\bf 598}, 509 (2001)
  [arXiv:hep-th/0010192].

\bibitem{Behrndt:1997fq}
  K.~Behrndt and W.~A.~Sabra,
  ``Static N = 2 black holes for quadratic prepotentials,''
  Phys.\ Lett.\  B {\bf 401}, 258 (1997)
  [arXiv:hep-th/9702010].

\bibitem{Sabra:1997dh}
  W.~A.~Sabra,
  ``Black holes in N = 2 supergravity theories and harmonic functions,''
  Nucl.\ Phys.\  B {\bf 510}, 247 (1998)
  [arXiv:hep-th/9704147].

\bibitem{Behrndt:1997ny}
  K.~Behrndt, D.~Lust and W.~A.~Sabra,
  ``Stationary solutions of N = 2 supergravity,''
  Nucl.\ Phys.\  B {\bf 510}, 264 (1998)
  [arXiv:hep-th/9705169].

\bibitem{Behrndt:1998eq}
  K.~Behrndt, G.~Lopes Cardoso, B.~de Wit, D.~Lust, T.~Mohaupt and W.~A.~Sabra,
  ``Higher-order black-hole solutions in N = 2 supergravity and Calabi-Yau
  string backgrounds,''
  Phys.\ Lett.\  B {\bf 429}, 289 (1998)
  [arXiv:hep-th/9801081].

\bibitem{Ferrara:1976iq}
  S.~Ferrara, J.~Scherk and B.~Zumino,
  ``Algebraic Properties Of Extended Supergravity Theories,''
  Nucl.\ Phys.\  B {\bf 121}, 393 (1977).

\bibitem{Rainich}
  G.~Rainich,
    ``Electrodynamics in the general relativity theory,''
    Trans.\ Am.\ Math.\ Soc. {\bf 27}, 106 (1925).

\bibitem{Emam:2009xj}
  M.~H.~Emam,
  Phys.\ Rev.\  D {\bf 79}, 085017 (2009)
  [arXiv:0904.1951 [hep-th]].

\bibitem{Emam:2004nc}
  M.~H.~Emam,
  ``Calibrated brane solutions of M-theory,'' (2004)
  [arXiv:hep-th/0410100].

\bibitem{Wald:1984rg}
  R.~M.~Wald,
  ``General Relativity,''
{\it  Chicago, Usa: Univ. Pr. (1984) 491p}.

\bibitem{Nakahara:2003nw}
  M.~Nakahara,
  ``Geometry, topology and physics,''
    {\it  Boca Raton, USA: Taylor and Francis (2003) 573 p}.

\bibitem{Joyce}
    D.~Joyce. ``{Lectures on {C}alabi-{Y}au and special {L}agrangian geometry}'' (2001) [arXiv:math.DG/0108088].

\bibitem{Berger}
    M.~Berger.
    ``{Sur les groupes d'holonomie homog\`{e}ne des vari\'{e}t\'{e}s \`{a} connexion affine et des vari\'{e}t\'{e}s riemanniennes}''.
    Bull. Soc. Math. France, 225--238 (1955).

\bibitem{hep-th/0102152}
  N.~Riazi,
  ``Geometry and toplogy of solitons,''
  [arXiv:hep-th/0102152].

\bibitem{hep-th/9605032}
  L.~Andrianopoli, M.~Bertolini, A.~Ceresole, R.~D'Auria, S.~Ferrara, P.~Fre and T.~Magri,
  ``N = 2 supergravity and N = 2 super Yang-Mills theory on general scalar
  manifolds: Symplectic covariance, gaugings and the momentum map,''
  J.\ Geom.\ Phys.\  {\bf 23}, 111 (1997)
  [arXiv:hep-th/9605032].

\bibitem{Candelas:1985hv}
  P.~Candelas, G.~T.~Horowitz, A.~Strominger and E.~Witten,
  ``Superstring Phenomenology,''. {\it Presented at Symp. for Anomalies, Geometry and Topology, Argonne, IL, Mar 28-30, 1985 and at 4th Marcel Grossmann Conf. on General Relativity, Rome, Italy, Jun 17-21, 1985.
    Published in ANL Symp.Anomalies 1985:377 (QC20:S96:1985) Also in
    DPF Conf.1985:737 (QCD161:A6:1985) Also in Grossman Meeting
    1985:227 (QC6:M3:1985)}

\bibitem{hep-th/9506150}
  G.~Papadopoulos and P.~K.~Townsend,
  ``Compactification of D = 11 supergravity on spaces of exceptional holonomy,''
  Phys.\ Lett.\  B {\bf 357}, 300 (1995)
  [arXiv:hep-th/9506150].

\bibitem{Ferrara:1991na}
  S.~Ferrara,
  ``Calabi-Yau Moduli Space, Special Geometry And Mirror Symmetry,''
  Mod.\ Phys.\ Lett.\  A {\bf 6}, 2175 (1991).

\bibitem{Candelas:1989bb}
  P.~Candelas and X.~C.~de la Ossa,
  ``Moduli space of Calabi-Yau manifolds,''
    {\it Prepared for XIII International School of Theoretical
    Physics: The Standard Model and Beyond, Szczyrk, Poland, 19-26
    (1989)}. Nuc.\ Phys. B {\bf 355} 455 (1991).

\bibitem{hep-th/9508001}
  H.~Suzuki,
  ``Calabi-Yau compactification of type IIB string and a mass formula of the
  extreme black holes,''
  Mod.\ Phys.\ Lett.\  A {\bf 11}, 623 (1996)
  [arXiv:hep-th/9508001].

\bibitem{Craps:1997gp}
  B.~Craps, F.~Roose, W.~Troost and A.~Van Proeyen,
  ``What is special Kaehler geometry?,''
  Nucl.\ Phys.\  B {\bf 503}, 565 (1997)
  [arXiv:hep-th/9703082].

\bibitem{hep-th/0203247}
  J.~Garcia-Bellido and R.~Rabadan,
  ``Complex structure moduli stability in toroidal compactifications,''
  JHEP {\bf 0205}, 042 (2002)
  [arXiv:hep-th/0203247].

\bibitem{hep-th/9512043}
  P.~Fre,
  ``Lectures on Special Kahler Geometry and Electric--Magnetic Duality Rotations,''
  Nucl.\ Phys.\ Proc.\ Suppl.\  {\bf 45BC}, 59 (1996)
  [arXiv:hep-th/9512043].

\bibitem{Cecotti:1988qn}
  S.~Cecotti, S.~Ferrara and L.~Girardello,
  ``Geometry of Type II Superstrings and the Moduli of Superconformal Field Theories,''
  Int.\ J.\ Mod.\ Phys.\  A {\bf 4}, 2475 (1989).

\bibitem{Gutperle:2000sb}
  M.~Gutperle and M.~Spalinski,
  ``Supergravity instantons and the universal hypermultiplet,''
  JHEP {\bf 0006}, 037 (2000)
  [arXiv:hep-th/0005068].

\bibitem{De Wit:1985qn}
  B.~De Wit, P.~Fayet and P.~Van Nieuwenhuizen,
  ``Supersymmetry And Supergravity '84. Proceedings, Spring School, Trieste,
  Italy, April 4-14, 1984,''
{\it  Singapore, Singapore: World Scientific} 469p (1984).

\bibitem{hep-th/9812049}
  F.~Denef,
  ``Attractors at weak gravity,''
  Nucl.\ Phys.\  B {\bf 547}, 201 (1999)
  [arXiv:hep-th/9812049].

\end{thebibliography}
\end{document}